\PassOptionsToPackage{desactivate}{linenoaa} 
\documentclass{aa}  

\usepackage{graphicx}
\usepackage{txfonts}

\usepackage{amsmath}
\usepackage{physics}

\usepackage[dvipsnames]{xcolor}
\usepackage{hyperref}
\hypersetup{colorlinks=true, linkcolor=blue, urlcolor=RedViolet, citecolor=blue}
\usepackage{natbib}
\usepackage{orcidlink}

\usepackage{dblfloatfix} 

\usepackage{upgreek}

\usepackage{xspace}
\newcommand{\sri}{Sr\,\textsc{i}\xspace}
\newcommand{\srii}{Sr\,\textsc{ii}\xspace}
\newcommand{\sriii}{Sr\,\textsc{iii}\xspace}
\newcommand{\sriv}{Sr\,\textsc{iv}\xspace}
\newcommand{\srv}{Sr\,\textsc{v}\xspace}
\newcommand{\hei}{He\,\textsc{i}\xspace}
\newcommand{\rproc}{\emph{r}-process\xspace}
\begin{document}

   \title{
   Strontium and helium in the kilonova AT2017gfo: Origin of the\\ $1\upmu$m feature constrained via NLTE calculations
   }


   \author{Aayush Arya
          \inst{1,2}
          \thanks{Corresponding author: \textcolor{RedViolet}{ aayush.arya@nbi.ku.dk}}\orcidlink{0000-0001-6782-407X} \and
          Rasmus Damgaard\inst{1,2}
          \orcidlink{0009-0000-5765-4601} \and
          Albert Sneppen\inst{1,2}
          \orcidlink{0000-0002-5460-6126}
           \and
          David J. Dougan\inst{3}\orcidlink{0009-0008-3312-7426} 
           \and
          Stuart A. Sim\inst{3}\orcidlink{0000-0002-9774-1192}
           \and
          Connor P. Ballance\inst{3}\orcidlink{0000-0003-1693-1793} 
          \and
          Darach Watson\inst{1,2}\orcidlink{0000-0002-4465-8264} 
          }

   \institute{Cosmic Dawn Center (DAWN), Denmark
        \and
        Niels Bohr Institute, University of Copenhagen, Rådmandsgade 64, Copenhagen N, 2200 Denmark
         \and
        Astrophysics Research Centre, Queen’s University Belfast, Belfast, BT7 1NN, Northern Ireland, United Kingdom
}

   \date{Received date; accepted date}

 
  \abstract{Mergers of neutron stars are believed to be one of the primary sites for the synthesis of the universe's heavy elements via the rapid neutron capture process. AT2017gfo, the kilonova following GW170817 provided the first direct spectroscopic evidence of the \rproc happening in the universe. A prominent line feature near $1\,\upmu$m in its spectrum was attributed to strontium -- a claim that has been independently recovered by several teams. However, in recent years it has been debated whether the feature arises instead from helium. Here, we present non--local thermodynamic equilibrium (NLTE) radiative transfer modelling of the observed kilonova spectra, including detailed radiation-matter interaction physics for both strontium and helium. We make use of freshly calculated strontium atomic data for e$^-$ impact collisions, photoionization, and recombination processes.
  Our strontium model self-consistently reproduces the temporal evolution of the $1\,\upmu$m feature at early times, with its absence at $0.92$\,days to its clear emergence at $1.17$\,days. This transition mimics LTE, because at early epochs ($t\lesssim 1.5$\,days) the radiation field dominates the ionization state of the ejecta over thermal and non-thermal electron collisions. We further test if helium can form the feature under the same plasma conditions. The helium mass required at $1.17$\,days is comparable to the total ejecta mass, while a few percent by mass of helium suffices at 4.4 days. On the other hand, the strength of the strontium lines decrease with time, and may require a radially stratified abundance to consistently produce the feature. We conclude that strontium is required to explain the onset of the feature at early times, but helium can contribute to, or even dominate the feature at later epochs. Finally, we demonstrate that in addition to the geometry and spatial confinement, the spectral lineshape is sensitively affected by the radial density profile and ionization structure of the ejecta. We find that helium confined to the polar ejecta can account for all of the absorption at 4.4 days, but it cannot produce the observed emission. Our results underscore the necessity of NLTE physics for accurately constraining the yields and spatial distribution of $r$-process elements in kilonovae.
  }
    


    \authorrunning{Arya et al.}

   \maketitle
%

\section{Introduction}

The origin of the elements of the periodic table remains one of the most pressing fundamental questions in physics \citep{Burbridge1957,Cameron1957}. Mergers involving a neutron star are believed to be among the primary sites of \rproc nucleosynthesis, producing half of the periodic table's elements heavier than iron. Decades after they were first postulated to be a site (\cite{1974ApJ...192L.145L,SymbalistySchramm1982}, see \cite{2019MetzgerReview} for a historical account), it was on August 17, 2017 when LIGO discovered a gravitational wave merger signal from the coalescence of two neutron stars \citep{Abbott2017}. Rapid follow-up observations by several dozens of teams successfully identified \citep{Coulter2017} and observed the associated kilonova, AT2017gfo \citep{2017Arcavi,2017Chornock,2017Cowperthwaite,Drout2017,Evans2017,Kasliwal2017Science, 2022KasliwalSpitzer, Kilpatrick2017,2017Nicholl,2017Pian,2017Smartt,2017Tanvir}.

Later, \cite{2019NaturWatson} identified the element strontium in the spectra of the kilonova. Since then, this claim has been independently reproduced several times \citep{2021DomotoRprocessFeatures,Gillanders2022photospheric,2023PognanNLTESpectra,Shingles2023,2023Vieira}. This event then marks the first direct spectroscopic confirmation of \rproc nucleosynthesis happening in the universe. These studies show that singly ionized strontium (\srii), which has a triplet of lines near $1\,\upmu$m, causes the blueshifted absorption trough between $700{-}1000$\,nm, which appears in the spectrum of the kilonova across many epochs. However, \cite{2022Perego} and \cite{Tarumi2023} considered whether this $1\,\upmu$m feature might come instead from the $1083.3$\,nm line of \hei. 
This is a contrasting interpretation, which deserves thorough reflection.

Helium is expected to be produced in neutron star mergers via two mechanisms: $\alpha$-decay of radioactive isotopes in case translead nuclei are synthesized, and via $\alpha$-rich freezeout. It has been known since long that helium can be produced in neutron star mergers via $\alpha$-rich freezeout \citep{2013FernandezMetzger}. In recent hydroynamical simulations incorporating detailed neutrino transport, it has become increasingly clear that this mechanism can operate in parts of the ejecta irradiated by the neutrino-driven wind, where the electron fraction ($Y_e$) and specific entropy ($s$) favour the freezeout of $\alpha$-particles and result in substantial helium production \citep{2022Perego,Sneppen2024Helium2,2025Jacobi,2025Bernuzzi}. 

\cite{2022Perego} pointed out that helium produced via $\alpha$-decay could appear in the NIR spectrum of the kilonova at the same location as the \srii triplet. However, they concluded that the trace helium masses in their simulated dynamical ejecta was insufficient to explain the full feature. \cite{Tarumi2023} presented an analysis of helium and strontium, and showed that in their models, $0.2\%$ of helium by mass could cause the $1\,\upmu$m feature while their \srii feature faded with time. In an independent analysis of helium, \cite{Sneppen2024Helium1} found that due to photoionization, the abundance of helium needed to explain the emergence of the feature at 1.17\,days post-merger exceeded the total ejecta mass in the line-forming region. They concluded that it is unlikely that helium can single-handedly explain the feature at all times. On the other hand, they found that small amounts of helium can contribute to the feature at later epochs (e.g.\ 4.4\,days). 

It is important to address which element comprises most of the feature for several reasons. 
In the element abundance distribution, strontium sits right after the first \rproc peak. Therefore, strontium should be produced in any \rproc nucleosynthesis event even if it is light \rproc dominated. The fact that it has a simple atomic structure with low level-density makes the individual lines of \srii very strong and something we expect to see if enough of it is present in the kilonova ejecta. Indeed, strontium is also the most established neutron-capture element identification, and if that were mistaken, it would strongly imply the need to revisit any other line identifications in the spectra of AT2017gfo.

On the other hand, helium is expected to be a byproduct of nucleosynthesis depending upon the ejecta conditions. The processes that produce helium in current hydrodynamical simulations depend on the lifetime of the hypermassive neutron star (HMNS) remnant that was produced by the merger. Therefore, a helium-dominated feature would reflect the ejecta conditions as well as the physics of the HMNS remnant, including the equation of state of very high density matter \citep{Sneppen2024Helium2}.

Computing synthetic spectra under non-local thermodynamic equilibrium (NLTE) conditions, as was done by \cite{Tarumi2023} and \cite{Sneppen2024Helium1}, depends on our knowledge of atomic data that determines the rate of different processes such as collisions, photoionization, and recombination. At the time of \cite{Tarumi2023}, these were not known for \srii except for A-values of bound-bound transitions. In kilonovae, an important source of ionization are the fast $\beta$-decay electrons that also power the lightcurve. \cite{Tarumi2023} assumed that these $\beta$ particles thermalize completely in the ejecta. In reality, only a fraction of their energy does, depending on the local density.
Also, observationally the $1\,\upmu$m feature in the kilonova spectrum transitions from being absent at $t=0.92$ day, to suddenly appearing at $t=1.17$ days since merger. There is a lot of information in the precise timing when this feature appears \citep{Sneppen2024EmergenceHourByHour} which was not considered by \cite{Tarumi2023}. This then left a need to revisit the calculations to address the strontium vs. helium question while self-consistently modelling both elements.

Most spectral line features, even in comparatively simple stellar spectra are blended, and involve contributions from multiple species. However, due to their quite distinct atomic structure and ionization potentials, strontium and helium will show different line behaviour under varying thermodynamic conditions and radiation fields. This makes it possible to consider limiting cases, and address which of these species dominates the line feature at different times.

In this paper, we will address the extent to which the feature is consistent with a strontium or helium interpretation. For this, we homogeneously model both elements while incorporating detailed radiation-matter interactions, with a full NLTE solution. The organization of the manuscript is as follows: In Section \ref{sect:sr-modelling-details}, we outline the kilonova ejecta model, the collisional-radiative model, and details of the spectrum calculations. In Section~\ref{sect:results-strontium}, the synthetic spectra are then compared to the observations. In Section~\ref{sect:NLTE-imp}, we emphasize the impact of NLTE physics on the plasma, the resulting spectrum, and its consequences for making inferences about element abundances. In Section~\ref{sect:geometry-inclination}, we consider the effect of geometry, 2D elemental distribution and observer inclination on the spectrum. We address the strontium vs.\ helium question in Section~\ref{sect:sr-vs-he}, and conclude with a summary in Section~\ref{sect:summary}. 


\section{Methodology}
\label{sect:sr-modelling-details}


Our calculation of the kilonova spectrum proceeds in the following fashion: First, we define the physical state of the ejecta (temperature, density, composition). The atoms in the ejecta experience a radiation field, which is coming from a sharply-defined photosphere that is emitting blackbody radiation. We divide the ejecta into 100 zones, on a grid of velocities ranging from $v=0.1c$ to $0.5c$. Next, we construct a collisional-radiative model and solve for the populations of the atomic energy levels, and the ionization state of the element in each of these zones. This model explicitly takes into account the rates of the different radiation-matter interaction processes such as line transitions, electron impact collisions, photoionization and electron-ion recombination. Then we collectively use this information to self-consistently compute the spectrum via radiation transport methods. 

 Unless stated otherwise, the kilonova is assumed to be spherically symmetric. In Section \ref{sect:geometry-inclination}, we will consider non-spherically symmetric ejecta and the consequence of their geometry on the spectrum. We assume that the ejecta is homologously expanding, such that at any point, the radius $r$ can be mapped to a velocity coordinate $v = r/t$, where $t$ is the time since merger.\footnote{Homologous expansion in neutron star merger ejecta starts to hold within a few hundred milliseconds since the merger for dynamical ejecta \citep{Neuweiler2023}, and a few seconds when including post-merger ejecta (e.g., \cite{groenewegen20252dendtoendmodelingkilonovae})} For the composition, our calculations assume that a single element (strontium or helium) with a certain abundance participates in the absorption and emission processes. In the following subsections, we describe the parameterization of our models, the processes included in the collisional-radiative model, and the radiation transport in detail.

\subsection{Temperature and radiation field}
\label{sect:radiation-field}

The temperature of the plasma directly sets the rate of different processes. The local radiation field drives photoexcitation and photoionization. In our models, the radiation field comes from a photosphere which is emitting as a single-temperature blackbody with temperature $T_{\rm{phot}}$. This is justified because at early epochs, the observed kilonova spectrum closely resembles a blackbody \citep{2023SneppenBlackbody}. Independent of the radiation temperature, one can define an electron temperature ($T_e$) of the plasma, which is a measure of the kinetic energy of the electrons and controls the rate of electron impact collisions and electron-ion recombination. In our calculations, we assume that $T_e = T_{\rm{phot}}$.

For a given epoch, we assume that the blackbody photosphere is situated in a shell of the kilonova ejecta having velocity $v_{\rm{phot}}$, determined from the best match to the Doppler shifts of the $1\,\upmu$m lines in the observed spectrum. The temperature of this blackbody was chosen by fitting a Planck function to the observed spectrum at the corresponding epoch, and correcting for Doppler boosts to the observer's line of sight, as outlined by \cite{2023SneppenBlackbody} and \cite{2025SadehPhotospheric}. Our adopted comoving-frame photospheric temperatures and velocities are listed in Table \ref{tab:temp-velocity}. 



Outside the photosphere, the mean intensity of this radiation field ($J_\nu$) is reduced compared to the Planck function ($B_\nu$) by geometric dilution $$J_\nu = W_{\rm{rel}}(v, v_{\rm{phot}}) B_\nu$$ 

\noindent where we define
$$ W_{\rm{rel}} = \frac{\gamma (1 - \mu_c')}{2} \left[1 + \frac{\beta(1+\mu_c')}{2}\right]$$ 
as the geometric dilution factor for a point in the ejecta at velocity coordinate $v$, corrected for Doppler transformations of the specific intensity and relativistic angle aberrations. Here, $ \mu_c' = (\mu_c - \beta)/(1-\mu_c\beta)$ is the cutoff angle cosine in the frame of the photosphere up to which rays from the photosphere can be seen at coordinate $v$, and $\mu_c = \sqrt{1 - v_{\rm{phot}}^2/v^2}$.

We provide a short derivation of this $W_{\rm{rel}}$ in Appendix~\ref{app:mean-intensity}.
 Note that in our time-independent treatment here, we have not taken into account effects due to the finite speed of light, which can be important during phases when the temperature is rapidly evolving. The above expression also assumes that $\beta$ is the same for all rays coming from the photosphere.


\begin{table}[!h]
\caption{The comoving-frame radiation temperature of the photosphere $T_{\rm{eff}}$, and spatial extent of the kilonova ejecta at different epochs. Note that under homologous expansion, $r_{\rm{min}} = v_{\rm {phot}}t$ and $r_{\rm{max}} = v_{\rm{max}}t$. 
}
\begin{center}
\begin{tabular}{@{}lccc@{}}
\hline\hline
Epoch (day) & $T_{\rm eff}$ (K) & 
$v_{\rm phot}$ ($c$)
& $v_{\rm max}$ ($c$) \\
\hline
0.92  & 5200 & 0.38$^{\textcolor{blue}{a}}$  & 0.50 \\
1.17  & 4900 & 0.295$^{\textcolor{blue}{b}}$  & 0.50 \\
1.43  & 4400 & 0.29   & 0.50 \\
2.42  & 3200 & 0.26   & 0.50 \\
3.41  & 2900 & 0.225  & 0.50 \\
4.40  & 2800 & 0.20   & 0.50 \\
\hline
\end{tabular}
\end{center}
\small

$^{\textcolor{blue}{a}}$In the absence of a line feature at $0.92$\,days, we estimate the $v_{\rm{phot}}$ from the luminosity distance using the observed spectral flux $F_{\lambda} = f(\beta)(v_{\rm{phot}}t/D_L)^2 B_\lambda$ and adopt a $v_{\rm{phot}}$ that gives a distance of $44\,$Mpc, similar to \cite{2023SneppenHubbleConstant}. 

$^{\textcolor{blue}{b}}$The photospheric velocity is poorly constrained for this epoch due to the lack of wavelength coverage at NIR wavelengths in the spectrum.
\label{tab:temp-velocity}
\end{table}

\subsection{Mass distribution of the ejecta}

\label{sect:density-profile}


In addition to the geometry of the ejecta, the spatial distribution of the mass, i.e. the density profile is important. The mass density profile $\rho(r)$ of the ejecta sets the density of free electrons ($n_e$) that the atoms present there can recombine with ($n_e \propto \rho$). Therefore, the spatial profile of $n_e$ will set the radial ionization structure of the ejecta. As time evolves, the observed line feature's blueshift becomes smaller and smaller, as the photosphere recedes inwards in velocity space. Therefore, at different epochs, the same absorption feature is probing different parts of the kilonova ejecta. A direct implication of this that, the $n_e(r)$ profile also sets both the spectral lineshape and the time evolution of the feature.

\begin{figure}[!h]
    \centering
    \includegraphics[width=0.49\textwidth]{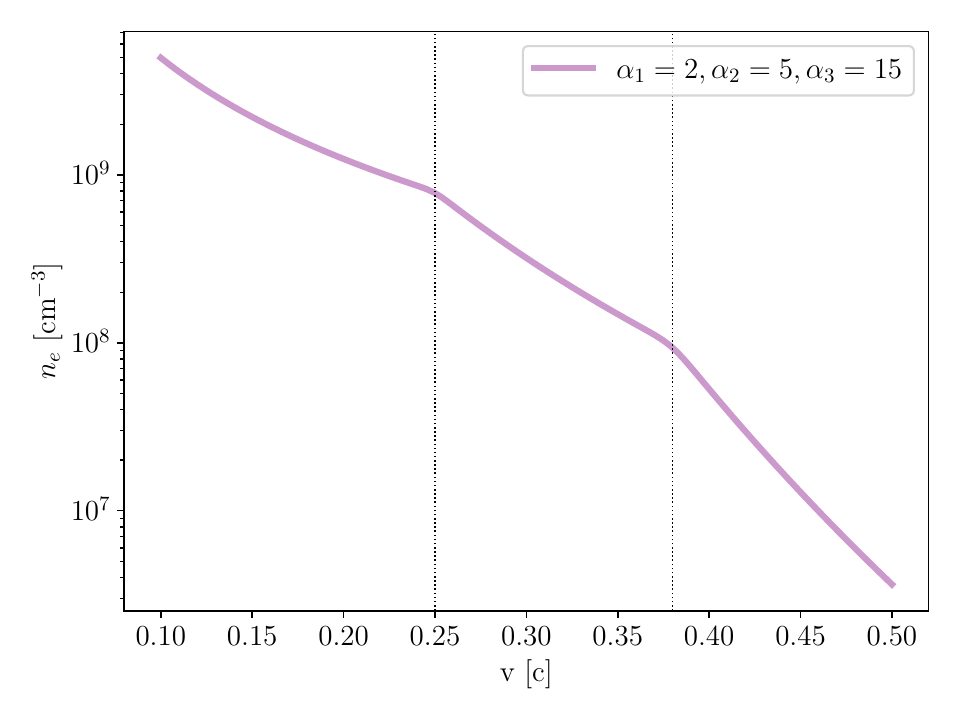}
    \caption{
    The density profile of mass and thus also the electron density $n_e$ is assumed to be a power law which is smoothly broken at $v=0.25$c and $v=0.38$c separating the power law slopes of $v^{-2}$, $v^{-5}$, and $v^{-15}$. We do not consider any ejecta mass at velocities $v<0.1c$. The numbers quoted are at $t=1$\,day.}
    \label{fig:broken-power-law}
\end{figure}

We assumed that the mass distribution of the ejecta follows a power law broken at multiple points, and our chosen density profile is shown in Fig.~\ref{fig:broken-power-law}. We chose a multiply broken power law instead of a single power law because varying slopes are expected from theoretical works. There is expected to be a larger concentration of mass at lower velocities $\lesssim 0.2c$ because when nuclei form, their $\approx 8$ MeV per nucleon binding energy is converted into a kinetic energy corresponding to $0.15{-}0.2c$. Mass at higher velocities is expected to come from dynamical and shock-heated ejecta, which can be a smaller fraction than post-merger ejecta. Indeed, in hydrodynamical simulations the  mass ejection results in flatter slopes in the inner ejecta, and steeper slopes in the outer ejecta. Current state-of-the-art simulations \citep[e.g.][]{Fujibayashi2023} have radial density slopes of  varying steepness throughout the ejecta profile.

    To have a power law that is broken with smooth transitions of the power law index (i.e., without kinks), we generalized the analytical form implemented in \texttt{astropy} within the module \texttt{SmoothlyBrokenPowerLaw1D} to multiple power law slopes. Our chosen profile then has the following analytical form $$ \rho = A \left( \frac{v}{v_{\mathrm{break,1}}} \right)^{-\alpha_1} \prod_{i=1}^{M} \left[ 1 + \left( \frac{v}{v_{\rm{break},i}} \right)^{1/\Delta_i} \right]^{-(\alpha_{i+1} - \alpha_i) \Delta_i}$$ 

\noindent where we adopted the slopes $\alpha_1 =2$, $\alpha_2 = 5$, $\alpha_3 = 15$ with breaks at  $v_{\rm{break,1}} = 0.25c$, $v_{\rm{break,2}} = 0.38c$ and sharpness of the break $\Delta_i=0.01$ for both of them. Note that there is no physical reason to prefer this parametric form over any other. A broken power law is merely a convenient function to use in that it allows varying the power law index in an interpretable way.   
    As we will describe in Section \ref{subsect:emergence-1um}, this choice of parameters allows us to reproduce the time at which the $1\,\upmu$m lines appear in the spectrum while providing a reasonable match to the observed spectral lineshape. A density profile that is too steep early on and does not have high enough electron densities at high velocities is unable to explain the most blueshifted absorption seen in the observed spectra.


    The kilonova ejecta must maintain charge neutrality. Therefore, the electron density is equal to the density of atoms, multiplied by the number of free electrons per atom ($f_e$), such that $$ n_e = f_e n_{\rm{A}}$$

\noindent where $n_A$ is the number density of atoms, if we take the mean mass number to be some $\expval{A}$. We chose $f_e$ to be a quantity that increases linearly from $2$ to $3$ from $v=0.1$ to $v=0.5$. As we will show later in Section \ref{subsect:ionization-balance}, the ionization structure of strontium suggests a mean ionization state similar to this, but we remain agnostic to the fact that other species could have a lower mean ionization (e.g.\ lanthanides due to their higher dielectronic recombination rates). 
    
    The normalization of the density profile is chosen such that the mass enclosed between $0.1c$ to $0.5c$ is equal to $0.04$M$_{\odot}$ while assuming a mean atomic mass of $\expval{A} m_p$, where as before $\expval{A}$ is 140. A lower $\expval{A}$, for a given number of free electrons per atom $f_e$ will increase the free electron densities $n_e$, and thus increase the recombination rates. Note that we have assumed the same electron density profile for the He model as for Sr. Throughout this article, elemental abundances will be quoted in terms of mass fractions of the total ejecta comprised by that element (e.g., $X_{\rm{Sr}}$).  

\subsection{Collisional-radiative modelling}

\begin{figure}[!h]
    \centering
    \includegraphics[width=0.45\textwidth]{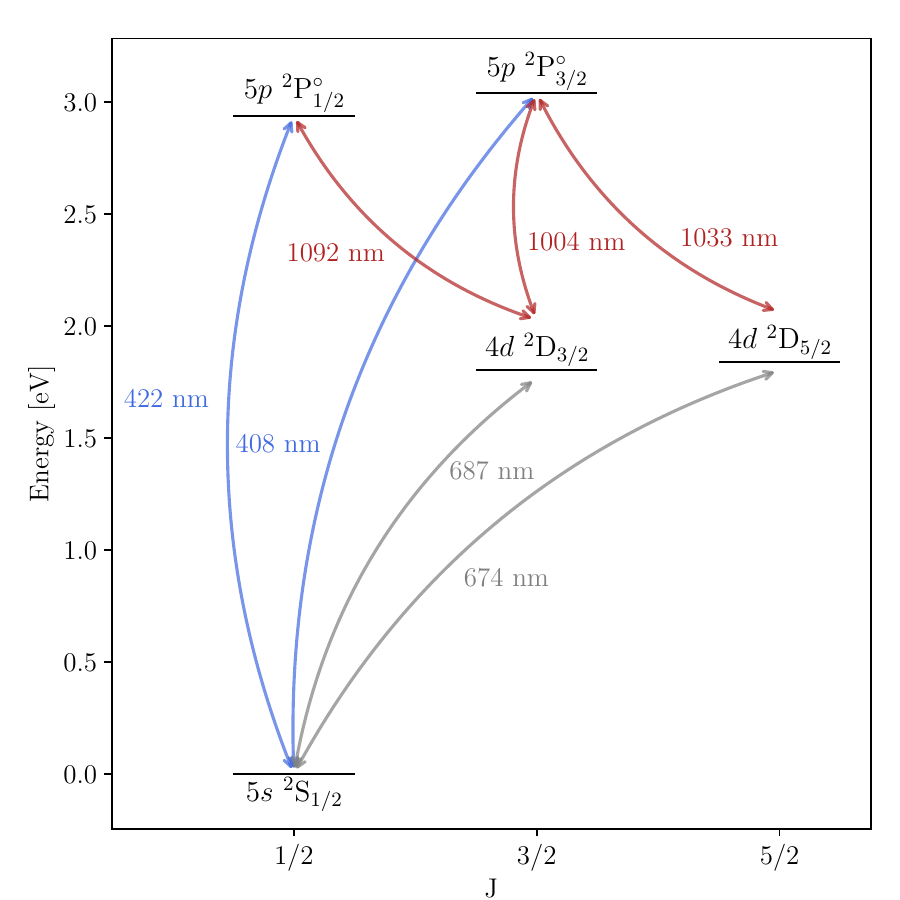}

    \caption{Grotrian diagram of \srii. The strong $400$\,nm doublet lines come from ground-state transitions, while the $1\,\upmu$m triplet responsible for the feature arises from the metastable levels.}
    \label{fig:grotrian-sr}
\end{figure}

\subsubsection{Strontium}

We use essentially the same collisional-radiative model as \cite{Tarumi2023}, with updated atomic data for strontium collision rates, photoionization cross-sections, and recombination rates. 
The strontium lines responsible for the $1\,\upmu$m feature particularly come from \srii. In Fig.~\ref{fig:grotrian-sr}, we show a Grotrian diagram of \srii with the important energy levels that we have included, and the bound-bound transitions that can occur. The strongest lines of strontium are at $408$\,nm and $422$\,nm, which come from transitions between the ground state $^2$S$_{1/2}$ and the excited states $^2$P$^{\circ}_{1/2}$ and $^2$P$^{\circ}_{3/2}$. Meanwhile, the lines at $1004$nm, $1033$nm, and $1092$\,nm (henceforth referred to as the 1\,$\upmu$m triplet) connect these excited states with two metastable levels $^2$D$_{3/2}$ and $^2$D$_{5/2}$. The doublet lines near $400$\,nm are much stronger than the $1\,\upmu$m triplet due to larger Einstein $A_{21}$ coefficients. 

 We have only included levels $\lesssim 3$\,eV, because there are no bound states of \srii up to another $\sim 3$\,eV above this, and levels that are above about 5\,eV will have negligible population at the temperatures relevant here for kilonova emission. For instance, at $4900\,$K the Boltzmann occupancy of the level at $5.91$\,eV is $\approx 10^{-7}$, and we expect it to have negligible contribution to absorption or emission features.
 
 For a complete ionization balance calculation, in our models we include all of the ionization stages of strontium from \sri to \srv. Among these, we include bound states of \srii, and the remaining ionization stages are represented as a single state alone. This is justified because neutral strontium (\sri) is expected to be a trace quantity even under local thermodynamic equilibrium (LTE), and the higher ionization stages of strontium (\sriii and above) do not have any low-lying bound states with optically allowed lines in the visible/NIR wavelengths.
 
 As we will show later, with the non-thermal ionization physics included here, higher ionization states are more abundant and neutral strontium is completely negligible. \sriv and \srv have ground-state forbidden lines at $1027.69\,$nm and $1203.35\,$nm respectively \citep{dougan2025strontiumiiiiiv}, but since due to their minuscule A-values ($A_{21} \approx 10^1\,$s$^{-1}$) those forbidden lines do not become optically thick, we do not expect them to contribute to the absorption feature. \cite{jerkstrand2025infraredspectralsignatureslight} included \sri to \sriv in their nebular phase NLTE calculations and did not find the $1027.69\,$nm forbidden line to contribute to a strong feature in emission either.

\subsubsection{Helium}

The collisional-radiative model of helium used in this work is the same as in \cite{Sneppen2024Helium1}, to which we refer the reader for further details. We summarize key aspects of it below.

The $1083.3$\,nm line of \hei arises from transitions between states of triplet (ortho) helium, namely 1s2s$^3$S $\leftrightarrow$ 1s2p$^3$P, whose lower level is 19.8\,eV above the ground state. In accordance to LS coupling selection rules, the transition from this lower level to the true ground state of \hei is strongly forbidden, and thus serves as a pseudo-ground state, separating the transitions happening within triplet (ortho) states of helium from those between singlet (para) states.

To reduce computational cost, fine structure sublevels were coarse-grained into one, by summing over the lower levels and averaging over the upper levels, weighted by the upper level degeneracy $g_u$. We included the same level transition channels for helium as we did for strontium.

\subsection{Level transition pathways}

\subsubsection{Radiative line transitions}

For the strontium and helium models, we include radiative line transitions between the bound states within \srii and \hei respectively. The atomic data including energy levels, statistical weights, line wavelengths, and $A$-values was taken from the NIST database \citep{Kramida2024}. These line transitions are driven by the blackbody radiation field of the photosphere described in Section \ref{sect:radiation-field}. 

\subsubsection{Photoionization}
Photoionization was included for \srii~$\to$~\sriii using freshly calculated level-resolved cross sections $\sigma(\nu)$ (D. J. Dougan, 2025, priv. comm.). The rates were evaluated from the raw cross sections as $$ \Gamma_{\rm{PI}} = \int_{\nu_{th}}^\infty \frac{4\pi J_\nu \sigma(\nu)}{h\nu}  d\nu $$ where $J_\nu = W_{\rm{rel}} B_\nu(T_{\rm{phot}})$ is the mean intensity at the point in the ejecta where the rate is being evaluated, $W_{\rm{rel}}$ is the geometric dilution factor at that point (see Section~\ref{sect:radiation-field}), and $\nu_{\rm{th}}$ is the threshold frequency corresponding to the ionization potential of the species from a given bound state.

For the helium model, as in \cite{Sneppen2024Helium1}, the level-resolved photoionization rates were taken from \citet{2010Nahar}.

\subsubsection{Electron impact collisions}

Excitation and de-excitation by thermal electron collisions in the ejecta was included for \srii and \hei. For \srii, we use the state-specific, temperature-dependent collision rates computed via R-Matrix methods by \cite{2024Mulholland}. These rates were not available to \cite{Tarumi2023}, who assumed the effective collision strengths were equal to 1. While this is a reasonable estimate for forbidden lines whose dimensionless collision strengths $\Omega_{ij} \approx 2-4$ \citep{2024Mulholland}, doing so underestimated the rates for dipole-allowed transitions for which \cite{2024Mulholland} find typical values of $\Omega_{ij}$ around $8-40$. 

We include bound-free ionization via thermal electron-impact collisions for \srii~$\to$~\sriii. We computed the cross-sections using a Disorted Wave Configuration Average (DW-CA) approach, as described by \cite{Pindzola1986}. 
%



For the helium model, only electron impact excitation is included (i.e., only bound-bound, not bound-free ionization), with thermally averaged rates from \citet{1987BerringtonKingston}.

\subsubsection{Non-thermal ionization}

The kilonova is powered by the radioactive decay of \rproc nuclei whose products thermalize within the ejecta and provide the energy that gets radiated as light. The primary decay channel for most of the isotopes is $\beta$-decay, with a minority contribution coming from $\alpha$-decay and fission fragments. The electrons that come from these $\beta^-$ decays have energies on the order of hundreds of keV to MeV, which well exceeds the ionization energies of bound electrons in atoms. These fast electrons can easily ionize even higher ionization stages of strontium such as Sr \textsc{iii}, \textsc{iv} and \textsc{v}, each of which have ionization energies in excess of 40\,eV. The non-thermal electrons could therefore to a great extent control the ionization state of the ejecta.

Similar to \cite{Tarumi2023}, we included the ionization by these non-thermal $\beta$-decay electrons, with a corresponding rate given as: 
$$ R_\mathrm{non-thermal} = \frac{\dot{q}_{\rm{dep}}(t) \expval{A} m_p}{w_i}$$
where $\dot{q}_{\rm{dep}}$ is the net heating rate per gram that thermalizes, $\expval{A}$ is the mean mass number of the \rproc ejecta, $m_\mathrm{p}$ is the mass of a proton, and $w_i$ is the work function for species $i$. Physically, $w_i$ is the amount of energy a $\beta$-decay electron loses until it manages to ionize an atom from ionization stage $i \to i+1$ (e.g.\ converting \srii $\to$ \sriii). A lower value of $w_i$ thus corresponds to a higher ionization efficiency.

  Here, we adopted an analytic prescription for the radioactive energy deposition due to $\beta$-decay $$\dot{q}_{\rm{dep}}(t) = f_{\beta}(t)\dot{Q}_{\beta,0} t_d^{-1.3}$$ where $f_\beta(t)$ is the thermalization efficiency of the $\beta$-decay products and $t_d$ is the time since merger in days \citep{2016Barnes,2019ApJ...876..128K}. We took $\dot{Q}_{\beta,0} = 10^{10}$\,erg\,s$^{-1}$\,g$^{-1}$ at $t_d=1$\,day \citep{2019ApJ...876..128K}, and assumed $\expval{A}= 140$, which is expected for a robust \rproc. 
We adopted a thermalization efficiency for $\beta$-decay products \citep{2016Barnes,2019ApJ...876..128K}
$$ f_{\beta}(t) = p_e (1+t/t_e)^{-n}$$

\noindent with the fraction of energy partitioned into fast electrons $p_e = 0.2$, the temporal index $n=1.5$, and

$$ t_e \approx 15 \left(  \frac{\eta M_{\rm{ej}}}{0.01M_\odot}\right)^{2/3} \left( \frac{v_{\rm{max}}}{0.2c} \right)^{-2} \ \rm{days}$$

\noindent
where $t_e$ is the timescale with which thermalization of the $\beta$-decay electrons in the ejecta becomes inefficient. For our fiducial model, $M_{\rm{ej}}=0.04 M_\odot$, and $v_{\rm{max}} = 0.5c$. The dimensionless parameter $\eta$ varies throughout the radius our ejecta, and is used to account for the dependence of the timescale on the local density. \cite{2019ApJ...876..128K} derived their analytical expressions such that $\eta=1$ throughout for a uniform density sphere, but varies at each radius for non-uniform density profiles such as the broken power law adopted here. Physically, in the ejecta at radius $r$, the parameter $\eta$ is the ratio of the density $\rho(r)$ for the chosen radial profile, with the density of a uniform sphere of radius $R = v_{\rm{max}}t$ and mass $M_{\rm{ej}}$. We therefore account for this density dependence of $t_e$ in a given zone via the parameter $\eta$.

\begin{figure*}[!ht]
    \centering
    \includegraphics[width=0.95\linewidth]{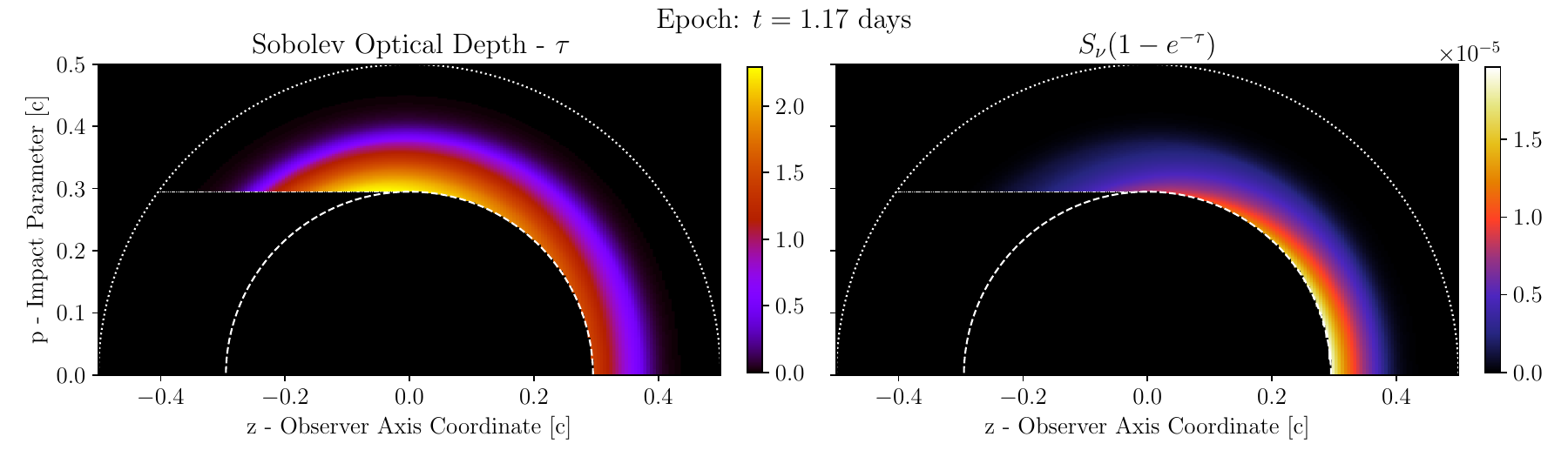}
    \caption{\textbf{Left}: The optical depth due to the $1033$\,nm line of \srii, in different parts of the ejecta. \textbf{Right}: The source function term in different parts of the ejecta showing emissivity from the same line. We note that even from a spherically symmetric ejecta, due to relativistic effects the observer frame contributions to these quantities from different parts of the ejecta are not uniform. While the source function is forward boosted, the optical depth has the opposite behavior. The dashed inner circle marks the photosphere ($r_{\rm{phot}} = v_{\rm{phot}}t$) while the dotted outer circle marks $v_{\rm{max}}t$. The part of the ejecta with $z<0$ and $p<v_{\rm{phot}}t$ is occulted out from the observer. We note that we have not included time-dependent effects due to the finite speed of light \citep{McNeill2025}.}
    \label{fig:source-function}
\end{figure*}

Note that, similar to \cite{Tarumi2023}, we have ignored thermalization of $\gamma$-rays emitted during the $\beta$-decay. For a uniform density ejecta, $\gamma$-rays are expected to be relevant only up to the peak of the optical lightcurve \citep{2016Barnes}, and their thermalization efficiency drops exponentially with the optical depth, $f_\gamma = 1- e^{-\tau_\gamma}$ \citep{2016Barnes,2019ApJ...876..128K}. Except for the earliest epochs where comparison can be made with the spectrum ($t\sim 1$ day), the $\gamma$-rays are essentially free-streaming. 

When relevant, $\gamma$-rays will either Compton scatter off electrons in the plasma, or cause photoionization of a bound electron. In principle, these Compton and photoelectrons should deposit their energy in the same way as $\beta$-particles. Therefore, our non-inclusion of $\gamma$-ray thermalization can be absorbed as an uncertainty in total radioactive heating rate, which we will discuss in Section \ref{sect:sr-vs-he}. For non-uniform density profiles, proper $\gamma$-ray thermalization requires a full treatment of transport as per the radiative transfer equation. It is known that deviations from analytic prescriptions can be substantial for $\gamma$-rays (see \citealt{Shingles2023}, Fig.~8). 

Meanwhile, for ionization by $\beta$-particles, we assume the same ``work function" $w_i$ as \cite{Tarumi2023}: $w_{\rm{\sri}} = 124$\,eV, $w_{\rm{\srii}} = 272$\,eV, $w_{\rm{\sriii}} = 444$\,eV, and $w_{\rm{\sriv}} =608$\,eV. The work function $w_i$ assumed for helium species are $w_{\rm{He\ I}}=593$\,eV and $w_{\rm{He\ II}}= 3076$\,eV. Note that in reality, $w_i$ depends on the composition such that a higher number of free electrons per atom increases $w_i$ as more energy then goes into heating instead of ionization. For a high mean ionization of the plasma, $w_i$ for each of these ions can become twice the quoted values, which lowers non-thermal ionization rates, impeding a further increase in mean ionization state of the ejecta. 

We have not included electron-impact excitation by non-thermal electrons as their contribution is expected to be important mainly for high-lying states (e.g.\ \citealt{Shingles2020ARTIS}) and most collisions with non-thermal electrons are expected to produce either heating or ionization \citep{KozmaFransson1992}. 

\subsubsection{Recombination}
We include temperature-dependent recombination rates of \srii~$\to$~\textsc{i} from \cite{2025Singh}, and the rates for \sriii~$\to$~\textsc{ii}, \sriv~$\to$~\textsc{iii} and \srv~$\to$~\textsc{iv} from McElroy et~al.\ (in prep.)
. For the latter three processes, both radiative (RR) and dielectronic recombination (DR) were included. Except for \srv~$\to$~\textsc{iv}, the dielectronic contribution at kilonova temperatures is small and radiative recombination dominates. For \srv~$\to$~\textsc{iv} the DR contribution exceeds RR even at low temperatures, due to resonances near the threshold. 


This is a distinction from \cite{Tarumi2023}, who in the absence of available strontium recombination rates assumed hydrogenic recombination rates \citep{Bates1962}. This is a good approximation for \sriii $\to$ \srii and has the right order of magnitude, but would underestimate the rates for \srv $\to$ \sriv, which are an order of magnitude higher. 

For strontium, we distribute the recombined electron among the different bound states of \srii per the statistical weights of the levels. We find that a different choice has no effect on the level populations. For the levels considered here, recombination is not an important mechanism for populating the excited states, as rates for bound-bound transitions that (de)-populate these levels are higher.

 However, the same does not hold for helium. Given that the 1s2s$^3$S state of \hei is $19.8\,$eV above the true ground state, photoexcitation and collisions from lower-lying states are very inefficient at populating this level. The population in this triplet helium pseudo-ground state comes primarily via recombination from He \textsc{ii} $\to$ He \textsc{i}. For helium, as in \cite{Sneppen2024Helium1}, we included level-resolved recombination rates with both radiative and dielectronic contributions taken from \citep{2010Nahar}.

 We do not explicitly track photons emitted during radiative recombination. In principle, photons produced by recombination from \sriii$\to$~\srii could get reabsorbed by other \srii atoms in the ejecta. The magnitude of this effect depends on the levels the recombined electrons end up in. A rough estimate of the optical depth to photoionization
 suggests that the ground and $^2$D metastable states are optically thick to their own recombination emission, while the photons emitted during recombination to higher-lying states can more easily escape. Level-resolved recombination rates computed by McElroy et al. (in prep) suggest that $\approx 55\%$ of the recombined electrons would end up in these states. If we consider an extreme scenario that each of these recombination photons is immediately reabsorbed, this would effectively reduce recombination rates by a factor of $\sim 2$.
 
 Note that, due to the large velocity gradients in the kilonova ejecta, photons are continuously redshifting. Therefore, recombination from other species could also contribute to this photoionization, and conversely, a species other than strontium could absorb the recombination emission leaving \srii less affected. 

 \begin{figure*}[!ht]
    \centering
    \includegraphics[width=0.95\textwidth]{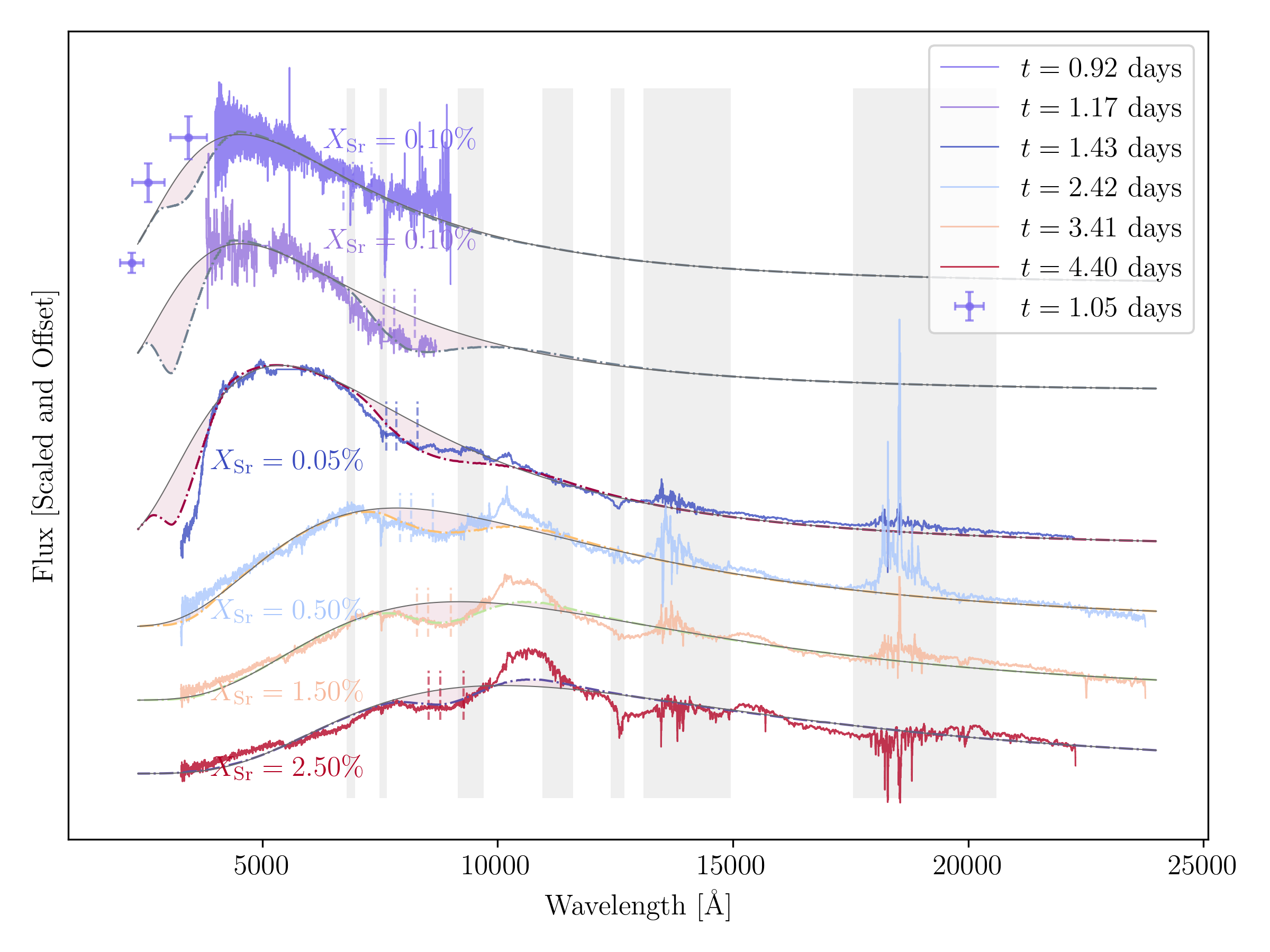}
    \caption{Observed spectra of AT2017gfo\protect\footnotemark\ \citep{2017Andreoni,2018Buckley,2017Pian,2017Smartt,Sneppen2024EmergenceHourByHour} and Swift UVOT photometry \citep{Evans2017} overlaid with the synthetic spectra from our NLTE strontium model, with the contribution of \srii to the spectrum highlighted in shaded pink. At 1.43 days, the $400$\,nm absorption arising from the resonance doublet is also simultaneously well-fit without additional tuning. The $1\,\upmu$m feature is not present at $0.92$\,days, and emerges within 6 hours at $1.17$\,days, which is well-recovered in the presented models. Shaded gray bands represent regions strongly affected by telluric absorption in the Earth's atmosphere.}
    \label{fig:fitted-strontium}
\end{figure*}


\subsection{Spectrum calculations}

We computed the spectrum using a ray-tracing code taking line-by-line optical depths of each transition. We use the level populations and ionization state of the ejecta obtained from the collisional-radiative model described above, to compute the Sobolev optical depth of each transition in each zone of the ejecta with $v \geq v_{\rm{phot}}$

$$ \tau^s_{ul} = \frac{\lambda^3 n_l \tilde{A}_{ul} }{8\pi} \left( \frac{g_u}{g_l} - \frac{n_u}{n_l} \right) t$$

where $\tilde{A}_{ul} = \tilde{\beta}A_{ul}$ is the effective $A$-value corrected by the Sobolev escape probability $\tilde{\beta} = (1-e^{-\tau})/\tau$, which accounts for self-absorption effects for optically thick lines. With this grid of $\tau$ in hand, we proceed with spectral line calculations. Note that while we include photoionization and recombination in our NLTE calculations for the level populations, the bound-free opacity and recombination emission are not included in the spectrum.

        We use our own line formation code based on the elementary supernova model \citep{JeffreyBranch1990}, but including the relativistic corrections of \citet{HutsemekersSurdej1990,Jeffrey1993}. 
        \footnotetext{Fetched from \url{https://github.com/Sneppen/Kilonova-analysis} under \texttt{Spectral Series of AT2017gfo}, see also \cite{Sneppen2024EmergenceHourByHour}.}
        The code\footnote{The code is available at \url{https://github.com/cartilage-ftw/kilonovae}} builds on developments by \citet{2019NoebauerSim,2023SneppenHubbleConstant} and includes some physics improvements which we describe below.

 In our present treatment, each line has its own source function, $S = \delta^3(\mu) S_{\rm{comov}}$, where the Doppler factor $\delta^3(\mu)$ transforms the comoving frame source function to the observer frame. 
$$ S_{\rm{comov}} = \frac{2h\nu_0^3}{c^2} \left[ \frac{g_u n_l}{g_ln_u} - 1 \right]^{-1}$$
We note that while our ejecta is spherically symmetric, in kilonovae these Doppler terms are large enough that observer-frame contributions from different parts of the ejecta become anisotropic (see Fig.~\ref{fig:source-function}). 

Our code does line formation by tracing specific intensity rays along the line-of-sight of the observer. A given specific intensity beam can see multiple resonances along its path, each of which attenuates the beam via absorption, or enhances it via scattering (the source function term). In our formalism, the inclusion of these processes happens while tracing the specific intensity beam $I_\nu$ as per equation (22) in \cite{JeffreyBranch1990}.

$$ I_\nu^{\rm{emit}} = I_\nu\exp \left(-\sum_{i=1}^N \tau_i\right) + \sum_{i=1}^NS_i(\nu)[1-e^{-\tau_i}] \exp \left(- \sum_{j=1}^{i-1}\tau_j \right) $$

\noindent where, the initial (continuum) specific intensity arising from the photospheric emission is $$ I_\nu = \begin{cases}
    B_\nu(T) & p \leq v_{\rm{phot}}t \\ 0 & \rm{otherwise}
\end{cases}$$ 
where $p$ is the impact parameter in Fig.~\ref{fig:source-function}. The $B_\nu(T)$ is evaluated directly in the observer frame, whose temperature $T$ comes from fitting a Planck function to the observed spectrum. Given this emergent specific intensity, the flux $F_\nu$ was computed by integrating over all impact parameters $p$ and azimuthal angles $\phi$, 

$$F_{\nu} = \int_{p=0}^{r_{max}}\int_{\phi=0}^{2\pi} I_\nu^{\rm{emit}} (p,\phi) p dp d\phi $$

In a homologously expanding ejecta, $r_{\rm{max}} = v_{\rm{max}} t$ where $t$ is the time elapsed since the explosion. Note that our code can model non-spherically symmetric ejecta structure, which we will consider in Section \ref{sect:geometry-inclination}.

\section{Strontium line features and their evolution}
\label{sect:results-strontium}
In Fig.~\ref{fig:fitted-strontium}, we show the synthetic spectra from our NLTE modelling overlaid on the observed spectra of AT2017gfo, and quote the abundance of strontium $X_{\rm{Sr}}$ needed to explain the $1\,\upmu$m feature at each epoch.

We find that a small amount of strontium is able to produce the observed $0.8{-}1\,\upmu$m absorption at early times. We emphasize the following key results from our NLTE radiative transfer work here: (i) In the observed spectrum, the absorption feature does not exist at $0.92$\,days, and emerges around $t=1.17$\,days \citep{Sneppen2024EmergenceHourByHour}. This evolution from the lack of a feature to the presence of one is well-reproduced in our models. (ii) Simultaneously matching the evolution of the feature, and the shape of the spectral line at $t=1.17$\,days can place powerful constraints on the kilonova's density structure. (iii) The $408\,$nm and $421\,$nm resonance doublet of \srii contributes to the absorption trough below $\lambda < 450$\,nm at $t=1.43$\,days. These lines remain optically thick at all epochs considered here, as is expected from the energy level structure. (iv) The $1\,\upmu$m emission is not well-reproduced in time-independent models. (v) The abundance of strontium required in our model to explain the $1\,\upmu$m feature increases with time. Note that our model here does not assume any radial stratification in the abundance of strontium.  



 We will revisit the last point in Section \ref{sect:sr-vs-he}, when we address the strontium vs. helium question. In what follows, we talk about the rest of the results in greater detail.  

\subsection{The emergence of the 1\,µm feature}

\label{subsect:emergence-1um}


The evolution of the $1\,\upmu$m feature from not forming in the $0.92$\,day spectrum, to appearing in the $1.17$\,days spectrum is something that emerges naturally even if with a small ($X_{\rm{Sr}} \sim 0.1\%$) amount of strontium is present in the kilonova ejecta. The reason for this evolution is that at $0.92$\,days, the radiation temperature is so high that due to photoionization, most strontium is present as \sriii or higher ionization stages and there is little \srii left to create the line absorption feature. But during the six hours between these two epochs, as the photosphere cools, the intensity of ionizing photons drops exponentially. Then, recombination is suddenly able to compete with ionization processes and a lot more \srii becomes available to form the observed $1\,\upmu$m spectral feature. This is a prediction that was made by \cite{Sneppen2024EmergenceHourByHour} in the LTE limit which we recover in our NLTE modelling.

\begin{figure}[!h]
    \centering
    \includegraphics[width=\linewidth]{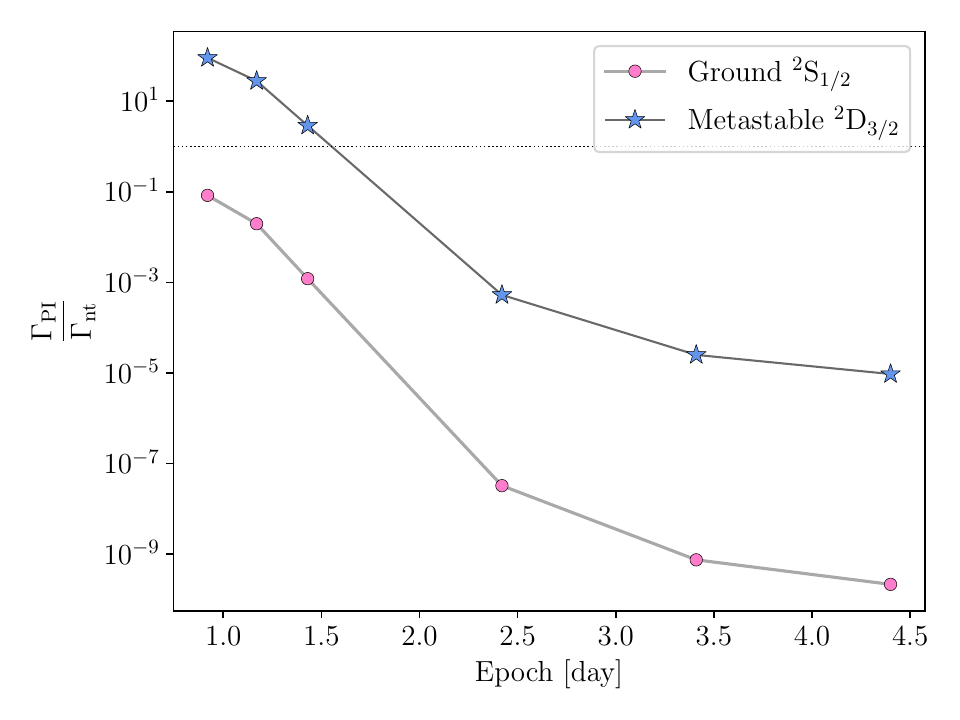}
    \caption{The ratio of photoionization to the competing non-thermal ionization by $\beta$-decay electrons. Photoionization of \srii is more important than non-thermal ionization at the earliest epochs, particularly from metastable and excited states. But with cooling radiation temperature, photoionization quickly becomes subdominant by $t \gtrsim 1.5$\,days. The range of photoionization rates across less than 4 days spans nine orders of magnitude.}
    \label{fig:ratio-photoion-nonthermal}
\end{figure}

This transition is controlled by the temperature of the radiation field, and the local electron density in the ejecta. During these times, the photoionization rate exceeds non-thermal ionization rate (see Fig.~\ref{fig:ratio-photoion-nonthermal}). 
Most of the photoionization proceeds through population in the $^2$D metastable and $^2$P excited states of \srii (see Fig.~\ref{fig:grotrian-sr} for an energy level diagram).  However, this changes after $t\gtrsim 1.5$\,days as the radiation field becomes less important and the ionization state of strontium starts being controlled more by non-thermal ionization. In Fig.~\ref{fig:ratio-photoion-nonthermal}, we show how the photoionization rate drops relative to non-thermal ionization by nine orders of magnitude between $0.92$\,days and $4.40$\,days. We will return to the implication of this for the subsequent evolution of the kilonova after $t\gtrsim 1.5$\,days in Section~\ref{subsect:recombination}.

The ionization threshold for \srii from the ground state is $11.03$\,eV and from the $^2$D metastable states is $\sim 9$\,eV. This means that this evolution of the spectral feature is relying on a mechanism that depends on the availability of photons with energy $\gtrsim 9$ eV. We note that while the true number of $\gtrsim 9$\,eV photons at $0.92$\,days is not known, the preceding and subsequent Swift UV photometry suggests that extrapolation of the observed kilonova spectra assuming a single temperature blackbody is a good approximation (see the first epoch in Fig.~\ref{fig:fitted-strontium}).

Our radiative transfer is time-independent, and our calculated optical depths rely on solving rate equations for the atomic level populations assuming that steady state holds. \cite{2022PognanSteadyState} studied the validity of steady state solution in detail, and found that it holds well when the relevant timescales for the microphysical processes are shorter than the evolutionary time. For the present discussion, the relevant microphysical processes are photoionization, non-thermal ionization, and recombination. We find that at the photosphere, the ratio of the rates $$ \frac{R_{\rm{ion}}}{R_{\rm{rec}}} = \frac{\Gamma_{\rm{PI}} + \Gamma_{\rm{nt}}}{\alpha_{i+1}n_e} $$

For \srii $\leftrightarrow$ \sriii, this ratio is greater than 1 at all epochs studied here, and therefore ionization timescales are shorter than recombination timescales. With this knowledge of the rate limiting factor, we can simply focus on the timescale for recombination, which is the amount of time a \sriii ion spends in the plasma finding an electron to recombine with
 $$ t_{\rm{rec}} = \frac{1}{\alpha_{i+1} n_e} $$
where $\alpha_{i+1}$ is the recombination rate coefficient for \sriii $\to$ \srii, and $n_e$ is the electron density in the relevant region of the ejecta. This recombination timescale was used in \cite{Sneppen2024Helium2} to constrain $n_e$, given the rapid observed appearance of the feature in absorption and emission on timescales of $\lesssim6$ and $\lesssim1$ hours, respectively.

At $t=0.92$\,days, at the photosphere the electron density while accounting for time delay effects is $n_e \simeq 1.35 \times 10^7$ cm$^{-3}$ , and the recombination timescale corresponding to $\alpha_{i+1} \approx 5 \times 10^{-12}$ cm$^3$/s is, $t_{\rm{rec}} \approx 4$ hours. As $t_{\rm{rec}}$ is much smaller than the evolutionary time, steady state should be valid, and our conclusions should hold. 

\subsection{The shape of the spectral line constraints the radial density profile}

One of the major unknowns in the modeling of the kilonova is its structure in terms of both geometry and radial density. The latter has thus far been poorly constrained. As we argued in Section \ref{sect:density-profile}, the number densities of the line-forming species as well as the local electron densities depend directly on the radial density structure. As a consequence, abundance inferences as well as the ionization structure of the ejecta are degenerate with it. It was shown by \cite{Sneppen2024Recombination}, if the electron density ($n_e$) at the photosphere alone is tuned as a free parameter, it is rather poorly constrained. The $n_e$ in that case can be varied by orders of magnitude while having a rather minor influence on the time at which the $1\,\upmu$m spectral feature emerges. 

However, the shape of the spectral line encodes information about the density structure that is complementary to that. In our multiply broken power law profile, if the breaks in the slopes are adopted at different velocities, it immediately affects the shape of the spectral feature that forms. Specifically, if we lower the electron density in the outer regions of the ejecta, there will be lesser \srii in those regions and the model will struggle to explain the most blueshifted absorption seen in the $t=1.17$\,days spectrum.

\begin{figure}[!h]
    \centering
    \includegraphics[width=\linewidth]{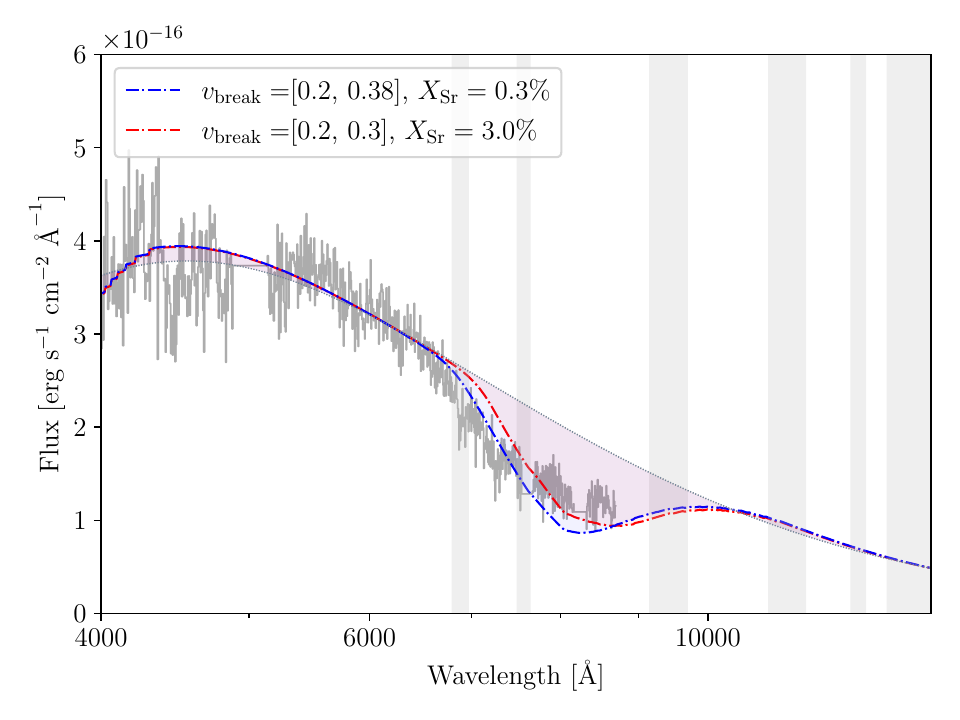}
    \caption{The dependence of the shape of the spectral line on the assumed ejecta radial density profile. For profiles where a steep power law is enforced too early (red curve), the model fails to produce the most blueshifted absorption seen in the $t=1.17$\,days spectrum. Therefore, fitting the shape of the spectral line can provide powerful constraints on the ejecta's density structure.}
    \label{fig:density-lineshape}
\end{figure}

We show this in Fig.~\ref{fig:density-lineshape}, where we calculated the spectral lineshape for two different density profiles, both with the same three power law slopes $n_e \propto v^{-\alpha}$, with $\alpha \in \{2,5,15\}$ separated at different $v_{\rm{break}}$ velocities. In one case, the steeper $v^{-15}$ slope is enforced already at $v=0.3c$ (red curve), while in the other, it is not placed until $v=0.38c$ (blue curve). In the case when the steep slope is enforced earlier, it can be clearly seen that, the most blueshifted absorption does not form even after adjusting with a higher $X_{\rm{Sr}}$ to compensate the reduced $n_{\rm{Sr}}$. This is because due to lower electron densities at those higher velocities, recombination from \sriii~$\to$~\textsc{ii} struggles to compete with ionization processes.

This shows that the spectral line is sensitive to conditions not just at the photosphere, but also in the entire line-forming region. We remind the reader that the electron density and mass density are tightly related to satisfy charge neutrality. Therefore, this implies that the mass in the kilonova ejecta at $v\gtrsim 0.3c$ in the ejecta of AT2017gfo cannot be smaller than a certain limit. Note that, this mass also cannot be larger than a threshold beyond which the feature cannot be suppressed in the $t=0.92$\,day spectrum. A very high mass (and thus electron density, $n_e$) correspondingly requires a high $n_{\rm{Sr}}$ to maintain charge neutrality, which will lead to a large optical depth of the line even at $0.92$ days. \cite{Sneppen2024Recombination} only varied the $n_e$ at the photosphere as a free parameter, keeping $n_{\rm{Sr}}$ fixed.

The density profile we adopted here was the best empirically matching one, while simultaneously accounting for all of the observations. A quantitative ``best fit" density profile can in principle be found by matching the observed spectral lineshape while still satisfying the above-mentioned observations. However, performing such a fit while quantifying degeneracies requires expensive Monte Carlo sampling of the parameter space, which is beyond the scope of the present study.

This dependence due to the ionization is seen also when comparing NLTE vs LTE models, which we will return to in Section \ref{sect:NLTE-lineshape}.

 \begin{figure*}[!ht]
    \centering
    \includegraphics[width=0.95\textwidth]{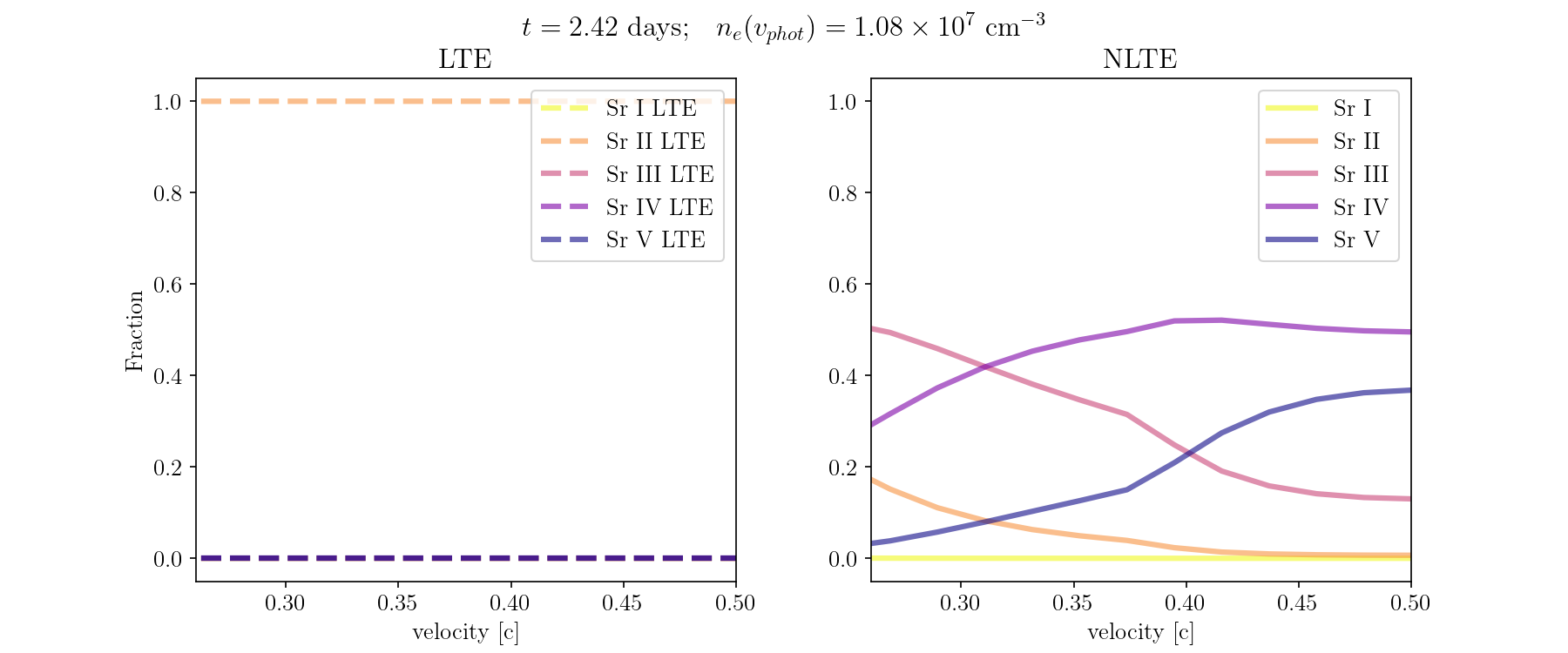}
    \caption{The ionization fraction of each atomic species in the kilonova ejecta, in different regions (velocity shells). \textbf{Left}: The expected ionization balance of different strontium species assuming LTE with temperature $T=3200$\,K, at $2.42$\,days and for the electron density profile shown in Fig.~\ref{fig:broken-power-law}. In LTE, \srii is expected to be the dominant ionization stage and, and no \sriii-\srv are expected. \textbf{Right}: The ionization fractions that result from the detailed NLTE calculations. Non-thermal ionization leads to a blended ionization structure, with ionization state increasing radially in the ejecta and \sriii-\srv dominating the composition. 
    The electron densities quoted are at the photosphere, which at $2.42$\,days lies at $v_{\rm{phot}}=0.26c$.}
    \label{fig:LTE_vs_NLTE_ionization}
\end{figure*}

\subsection{$400$\,nm absorption from the resonance doublet}

Our calculations predict that the \srii resonant doublet lines should cause absorption in the $1.43$\,day spectrum at $\lambda < 4500\,\rm{\AA}$, which is seen also observationally (see Fig.~\ref{fig:fitted-strontium}). As mentioned before, the doublet lines at $408$\,nm and $421$\,nm have larger $A$-values and greater population in the corresponding lower level than the $1\,\upmu$m triplet lines. This makes the doublet lines optically thicker than the $1\,\upmu$m triplet, across all epochs. Visually, it appears as if the $400$\,nm absorption has disppeared in the observed spectrum from $2.42$ days onwards. However, in our models the $\tau$ for these lines remains large, and absorption below $4500$\,$\rm{\AA}$ is predicted even at $t\gtrsim 2.4$\,days (see Fig.~\ref{fig:fitted-strontium}). Note that this absorption is weak in absolute terms because there is not much flux left at those wavelengths to absorb. But the lines still cause a fractional suppression of flux as per $\sim I_0 e^{-\tau}$, although this may be difficult to detect observationally. 

 We note that other authors \citep{Gillanders2022photospheric,2023Vieira} have attributed this UV-blue absorption trough to come primarily from Y\,\textsc{ii} and Zr\,\textsc{ii} lines instead of strontium. We believe that \srii, if present, should cause absorption in this region, if simultaneously the $1\,\upmu$m lines from \srii exist and are prominent in the spectrum.  The presence of strong lines from other elements in this wavelength region will not hamper the growth the \srii resonance doublet, unless the said strong lines are so close in wavelength that they really overlap with the atomic lineshape functions $\phi(\nu)$ of the doublet lines within the Doppler width (about $\sim 5$ km/s) and the Sobolev approximation breaks down. In that case, they would lower the escape probability of photons, reducing effective transition rates for the strontium doublet, making it have a weaker relative contribution. But indeed, the Y\,\textsc{ii} and Zr\,\textsc{ii} lines do not overlap this closely.

    One possibility is that for the lines of Y\,\textsc{ii} and Zr\,\textsc{ii} that are bluewards of the \srii doublet lines, a beam of photons that is redshifting while traversing through the ejecta will see a resonance with the bluer Y\,\textsc{ii} and Zr\,\textsc{ii} lines first, which will already absorb most of the photons from it. Then, the number of photons that \srii can remove becomes smaller. This is not because the \srii lines are not strong, but nonetheless the contribution they can have in forming the absorption trough becomes tiny.
 
    Another simplification true for our model is that it assumes a wavelength-independent photosphere. That is, the spatial depth at which the photosphere is located and the point at which the radiation starts to escape, is the same in our model for the $400$\,nm region as for the $1\,\upmu$m triplet. A higher opacity from substantially more bound-bound transitions in the bluer wavelengths could mean that this is not true in reality and the radiation may be escaping further out in the ejecta for bluer wavelengths than it does for the $1\,\upmu$m region. This change in the extent of the line forming region would affect the observed shape and strength of the resonance doublet and would mean that while \srii contributes to the observed absorption, it does not solely explain the entire trough.

\subsection{Modelling the emission}

Our models do not match the apparent strength of the emission part of the feature. This could in part be due to the fact that our line calculations assume the speed of light is infinite, which neglects light travel time effects, that have been shown to affect the strength of the emission for these lines \citep{Sneppen2024Recombination,Sneppen2024EmergenceHourByHour,McNeill2025}. Time-independent radiative transfer struggles with this in general, and the weaker strength of the emission is also seen in works with other codes such as \textsc{tardis} \citep[e.g.\ see the leave-one-out spectra in Fig.~7 of][]{2024Vieira}.

Physically, the light from different parts of the kilonova ejecta reaches the observer at different times, and at a given observer time (e.g.\ $t=1.43$\,days) there can be several hours delay between when photons arrive from certain parts of the ejecta. A consequence of this is that photons which get released around the $z\approx0$ plane (see Fig.~\ref{fig:source-function}) and contribute to the rest-wavelength emission, come from a portion of the ejecta that has different thermodynamic conditions than where the highest velocity absorption emission forms (around $z \sim v_{\rm{max}}t$). Due to the fact that the temperatures are rapidly evolving, at the same observer time, the emission would come from a region that experienced hotter temperatures than the region where the blueshifted absorption comes from.

We expect this to have an impact on the spectral lineshape, making the emission appear stronger. The reader is referred to \cite{Sneppen2024EmergenceHourByHour,McNeill2025} for a detailed discussion. In the future, if the inclusion of these effects does not suffice to explain the $1\,\upmu$m bump at $4.40$\,days, it could mean that the apparent emission does not come from scattered photons. It might instead be part of the continuum emitted by the photosphere, which deviates from the fitted blackbody we have adopted here. 


 Nonetheless, our main conclusions are robust to changes in the relative strength of the emission.






\section{Consequences of NLTE physics}

\label{sect:NLTE-imp}

\subsection{Ionization Balance: \srii contributes as a minority species}

\label{subsect:ionization-balance}

Before \cite{Tarumi2023} and \cite{2023PognanNLTESpectra} most studies of radiative transfer for kilonovae that included strontium had been done with the assumption of local thermodynamic equilibrium (LTE). 
At times when the radiation field dominates the ionizing processes ($t\lesssim 1.5$\,days), our results track the trends predicted by LTE, including major transitions in ionization state. After photoionization becomes sub-dominant to non-thermal ionization, however, the evolution is substantially different and LTE underestimates the mean ionization state. For instance, LTE predicts that at $t=2.42$\,days with the $3200$\,K temperature of the photosphere controlling the plasma state, the whole ejecta would uniformly possess \srii as the dominant ionization state comprising nearly $100\%$ of total strontium out to the highest velocity regions (see Fig.~\ref{fig:LTE_vs_NLTE_ionization}). However, similar to \cite{Tarumi2023,2025BrethauerIonization}, we find that accounting for non-thermal ionization results in a mix of multiple ionization states at the same time, with \srii to \srv
 coexisting across all epochs. In fact, \srii is a minority species throughout. The ionization structure is inverted, with the mean ionization state increasing radially outwards in the ejecta, with there being negligible \srii in the outermost ejecta. 


While non-thermal ionization is a general process affecting all species in the kilonova ejecta, other species could have a lower mean ionization state if their recombination rates are higher. 
Indeed, \cite{2021Hotokezaka} in their neodymium modelling found that Nd\,\textsc{ii} and \textsc{iii} were the dominant ionization stages in most epochs in the nebular phase. We attribute this difference to the two orders of magnitude higher dielectronic recombination rate for neodymium ($\alpha_{\rm{Nd\,\textsc{iii}}}\sim 10^{-10}\,$cm s$^{-1}$; \citealt{2021Hotokezaka}) than strontium ($\alpha_{\rm{\sriii}} \sim 10^{-12}\,$cm s$^{-1}$; McElroy et al. (in prep)).

Regardless, if it is true that singly ionized species are less abundant than doubly and higher ionized species for many of the \rproc elements as well, this may address a difficulty that current LTE radiative transfer work struggles with -- a very high opacity from bound-bound transitions at blue wavelengths, producing much more absorption than is observed \citep{Gillanders2022photospheric,2023Vieira,2024Vieira,vieira2025spectroscopicrprocessabundanceretrieval}. A lower abundance of singly ionized species would take away some of this opacity, which would be shifted to hard UV and soft X-ray wavelengths at which the dipole permitted transitions from the ground configurations of multiply-ionized species typically lie. This could be important for inferences of the lanthanide mass fraction of AT2017gfo, which is severely impacted by the high opacity of neutral and singly-ionized lanthanides \citep{GillandersLanthanideFrac2026}.
Indeed, with radiative transfer simulations using the \textsc{Sedona} code, \cite{2025BrethauerIonization} found that the higher mean ionization of the ejecta due to non-thermal ionization reduces line-blanketing in the optical bands, and affects the color evolution after the peak of the lightcurve. 

This ionization structure has serious implications including the time evolution of spectral features, occurrence of recombination episodes, and even for the shape and extent of P\,Cygni lines resulting from the same species. In what follows, we discuss these in further details.

\subsection{On recombination episodes in the plasma}
\label{subsect:recombination}

\begin{figure}[!h]
    \centering
    \includegraphics[width=\linewidth]{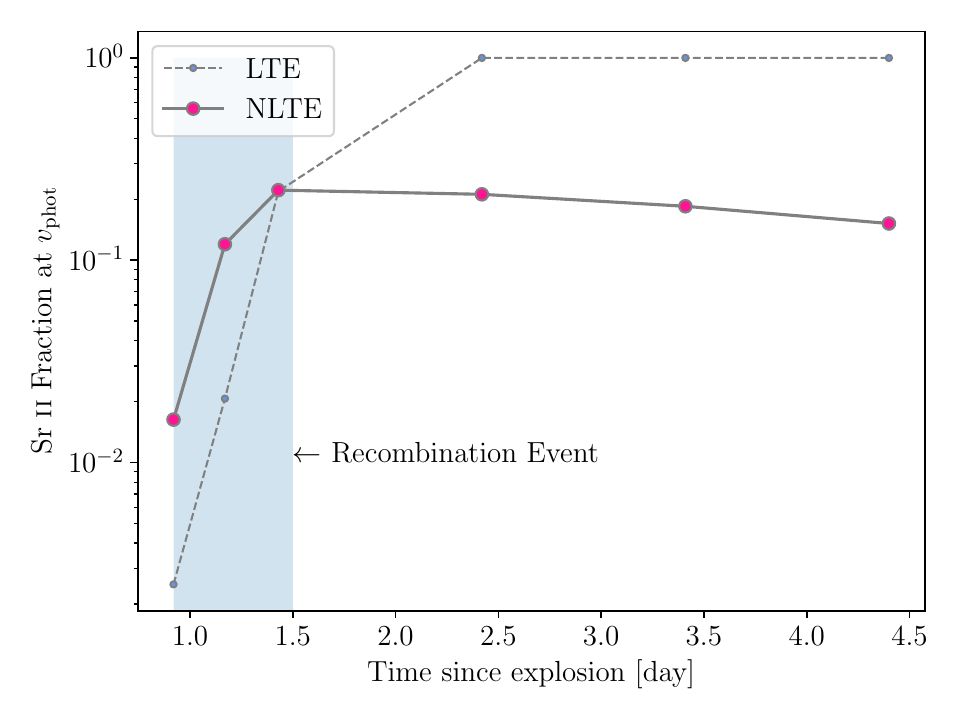}
    \caption{The fraction of strontium that is in the \srii ion state at the photosphere, across the epochs. The rapidly rising \srii around $t\approx 1$\,day with a recombination transition predicted by LTE also exists in the NLTE model, but ultimately stagnates into a multiple ionization-state structure. We quote the numbers at the photosphere $v_{\rm{phot}}$ (which is receding with epoch) not in a fixed velocity shell, where the mean ionization state keeps rising.}
    \label{fig:recombination-wave}
\end{figure}

As discussed in Section~\ref{subsect:emergence-1um}, the formation of the \srii absorption feature between $0.92$ and $1.17$\,days is understood as no \srii being available at the earlier epoch, followed by a rapid recombination episode throughout the ejecta suddenly leading to more \srii becoming available \citep{Sneppen2024EmergenceHourByHour}. The formation of this feature is consistent with the time at which such an episode is expected based on the observed blackbody radiation field setting the local ionization temperature \citep{Sneppen2024EmergenceHourByHour}. In Fig.~\ref{fig:recombination-wave}, we show that the fraction of \srii rises rapidly between $0.92$\,days and $1.43$\,days, resulting from recombination of \sriii $\to$ \srii. Therefore, the LTE expectation seems to be reproduced in our NLTE model. We address next why this is.

Physically, this recombination episode occurs not due to increasing recombination rates, but due to dropping ionization rates. If we consider the joint effect of dropping electron densities due to homologous expansion and the recession of the photosphere from $v_{\rm{phot}} = 0.38c$ at $0.92$\,days to $v_{\rm{phot}} = 0.295c$ at $1.17$\,days, the change in electron densities at $v_{\rm{phot}}$ is only around $\sim 40\%$ and the sensitivity of the recombination rate coefficient $\alpha_{i+1}$ due to changes in electron temperature over these timescales is even smaller. On the other hand, as we showed before, photoionization rates drop quickly by more than an order of magnitude over these six hours (see Fig.~\ref{fig:ratio-photoion-nonthermal}). 

We also note that the presence of this recombination episode is not due to a change in the local \emph{electron} temperature. The \sriii~$\to$~\srii recombination rates depend weakly upon temperature, varying only by $\sim 40\%$ percent between $3000$\,K and $5000$\,K (McElroy et al. (in prep)). We find that even if we artificially hold the electron temperature constant at $5500\,$K between these two epochs, it does not suppress the recombination. Therefore, the rapid emergence of $1\,\upmu$m feature with a recombination event should not be interpreted as cooling of the plasma's electron temperature or evidence for equilibration of matter and radiation. 

\cite{2021Hotokezaka,2022PognanSteadyState} predicted through first principles evolution of the thermodynamic structure of the kilonova ejecta that the electron temperature would rise with time in the nebular phase, from the beginning of their calculations at $t\approx 5-10$\,days. This results from the net effect of radioactive heat injection, line cooling, and adiabatic expansion of the ejecta. While the general principles apply, it remains to be seen if this holds also in the earlier photospheric phase of the kilonova.



 It is clear from Fig.~\ref{fig:recombination-wave} that when including non-thermal ionization, the recombination stagnates well before reaching the \srii fraction that is predicted by LTE, and, while still an appreciable fraction of the Sr, remains persistently well below the LTE value from $t=2.42$\,days to 4.40\,days. As time progresses, LTE would predict a second recombination event around 10\,days, where singly ionized species become neutrals. \cite{Sneppen2024EmergenceHourByHour} cautioned that due to the dominance of non-thermal ionization at these times, this may not be true.
 
 \cite{2025BrethauerIonization} considered the subsequent evolution of the ionization structure at later times. 
 As per standard prescriptions, radioactive energy generation declines as $t^{-1.3}$ and thermalization efficiency at times $t \gg t_e$ declines as $f_\beta \approx (t/t_e)^{-1.5}$, where $t_e$ is the timescale with which the thermalization of electrons becomes inefficient. Recombination rates in a given velocity shell drop continuously due to declining electron densities under homologous expansion, $n_e \propto t^{-3}$. Put together, if photoionization can be ignored, the ratio of the rates $$ \frac{R_{\rm{ion}}}{R_{\rm{rec}}}\propto  \frac{f_\beta(t) t^{-1.3}}{\alpha_{i+1} n_e(t)} \propto t^{0.2}$$
 However, in our models up to $t=4.40$\,days, in a given velocity shell we do not find the ionization to stop declining with time because (i) the photosphere is constantly receding to ejecta regions with lower velocities, which have longer $t_e$ timescales and so $f_\beta$ doesn't decline as fast in the relevant line-forming region, and (ii) for the assumed $M_{\rm{ej}}$ of $0.04\,$M$_\odot$ the $t_e$ is long enough in the inner ejecta that we does not reach the $t \gg t_e$ regime by $4.40$\,days. At the photosphere, this leads to an initial recombination episode around $t\approx 1$\,day and followed by an increase in mean ionization (see Fig.~\ref{fig:recombination-wave}). However, we do notice the effect of decreasing thermalization efficiency in the less dense outermost ejecta ($v \gtrsim 0.40c$) around $3.41$\,days already (see right panel of Fig.~\ref{fig:LTE_vs_NLTE_ionization}) whereby the mean ionization state starts to flatten out (instead of continuing to rise) because with decreasing densities in this region, the non-thermal ionization rates also become lower.

In either case, from this it seems unlikely that the ionization state would drop so strongly that neutrals could dominate. This then lends support to presence of multiply ionized species in the kilonova ejecta at later times such as claimed observationally for 
Te \textsc{iii} \citep{Hotokezaka2023Tellurium}, W \textsc{iii} \citep{Hotokezaka2022TungstenSelenium,2024McCann}, and (tentatively) Te \textsc{iv} \citep{2025MulhollandTellurium}.

\begin{figure}[!t]
    \centering
    \includegraphics[width=0.5\textwidth]{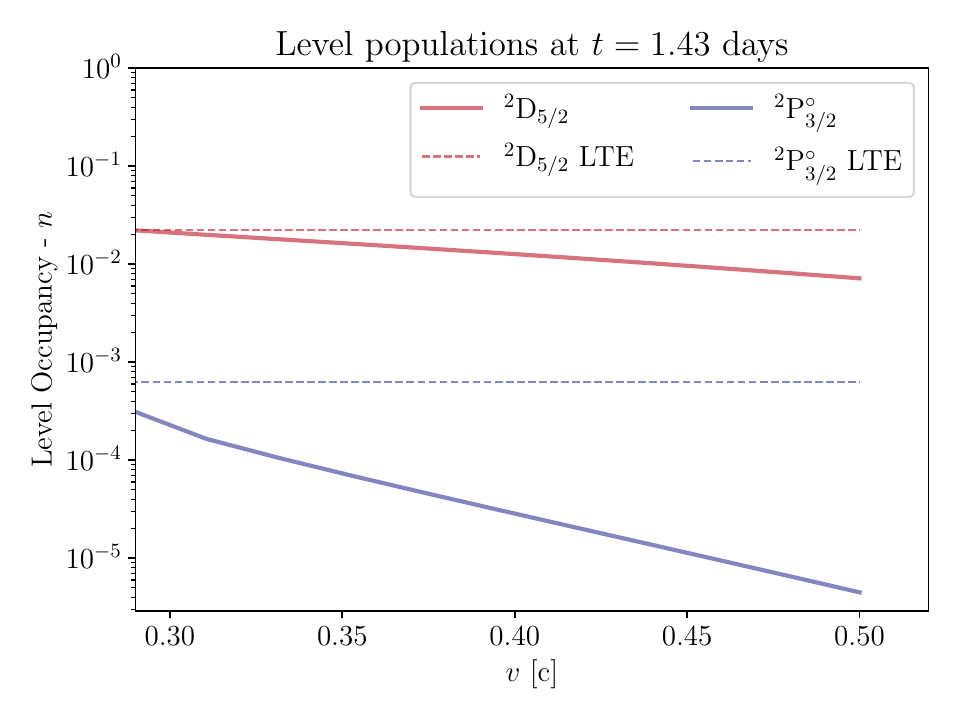}
    \caption{The level populations of a subset of the energy levels in \srii, across the line forming region at $t=1.43$\,days. At the photosphere $v_{\rm{phot}}=0.29$c, the population in the $^2$D$_{5/2}$ metastable level is close to LTE, while in the $^2$P$^\circ_{3/2}$ excited state population is half of LTE due to geometric dilution of the radiation field. The population in both decreases as one moves further out in the line-forming region. These two levels are the respective lower and upper levels of the $1033$\,nm line. The population in most of the line-forming region is not in LTE.}
    \label{fig:level-pops}
\end{figure}

\begin{figure}[!b]
    \centering
    \includegraphics[width=\linewidth]{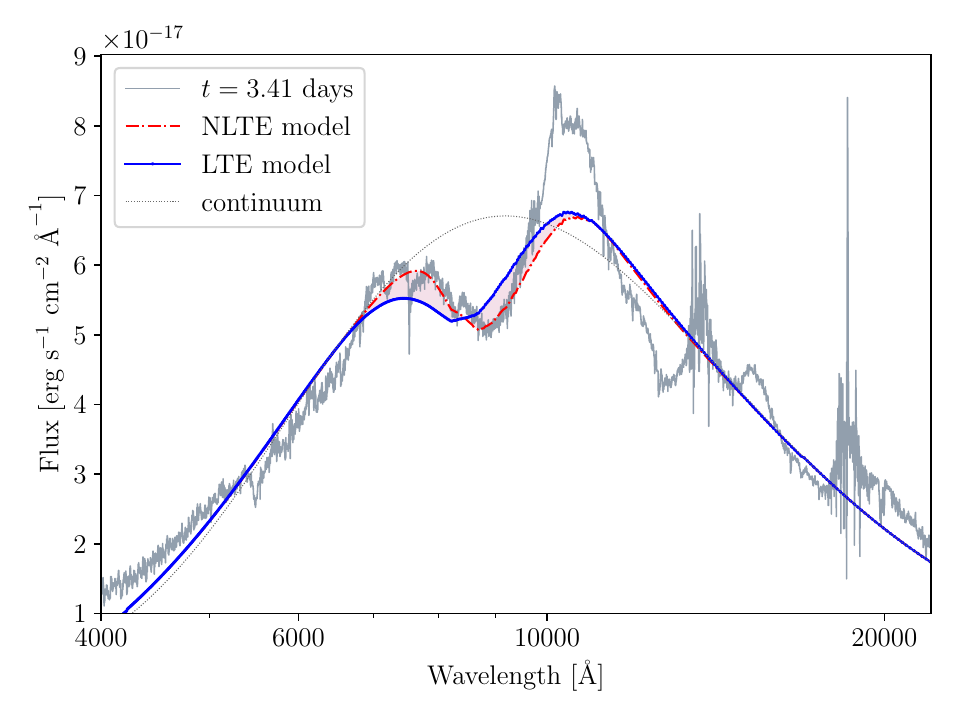}
    \caption{\srii $1\,\upmu$m lines from an LTE vs. NLTE model, assuming the same radial density profile. In LTE, as a lot of \srii exists even in the outermost ejecta, the absorption extends up to higher blueshifts. With NLTE ionization structure the bluest absorption is truncated earlier. Note that the LTE calculation assumed geometric diluted LTE populations for evaluating the source function.}
    \label{fig:lineshape-LTE-NLTE}
\end{figure}

\begin{figure*}[!b]
    \centering
    \includegraphics[width=0.49\textwidth]{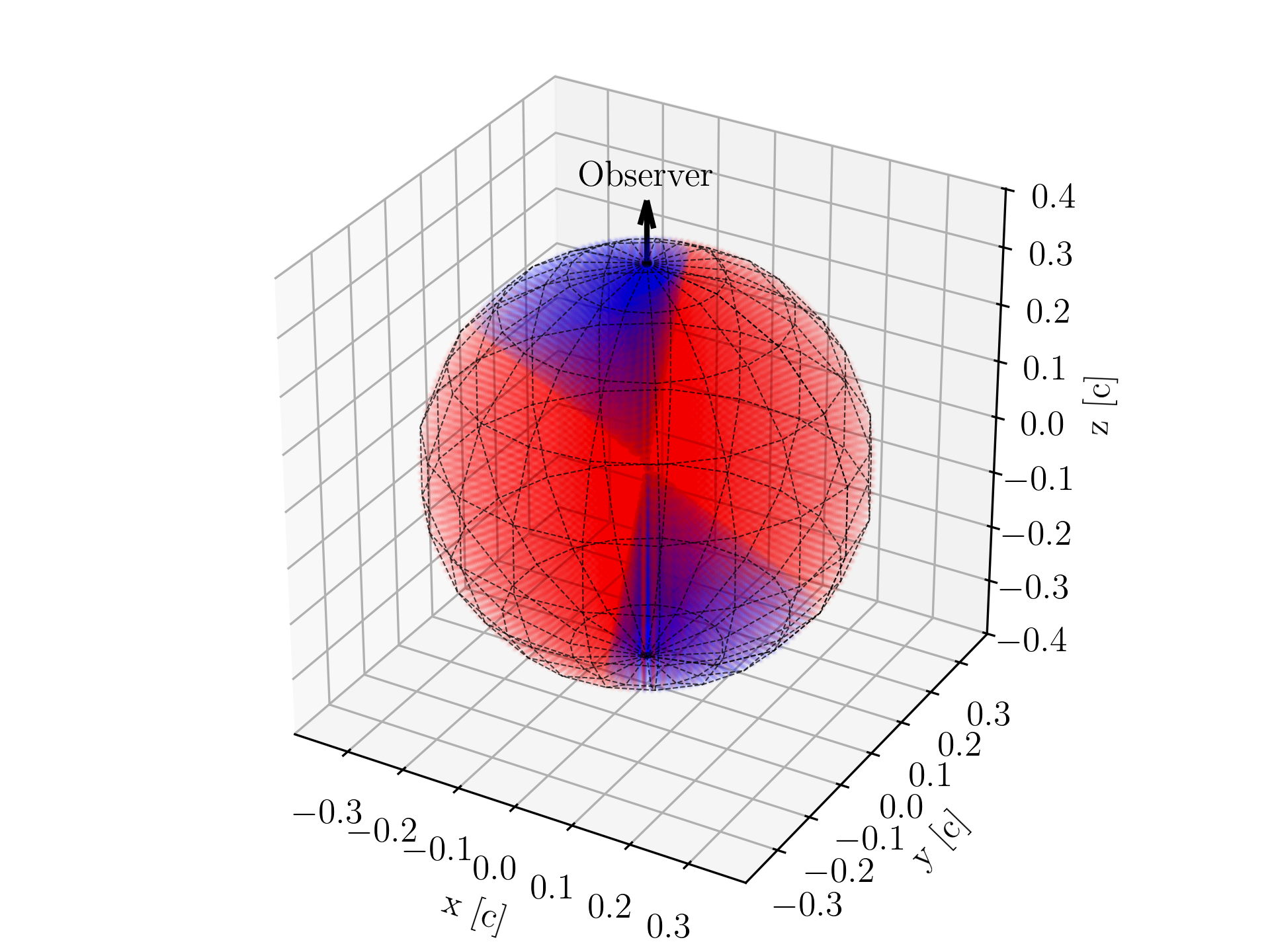}
    \includegraphics[width=0.49\textwidth]{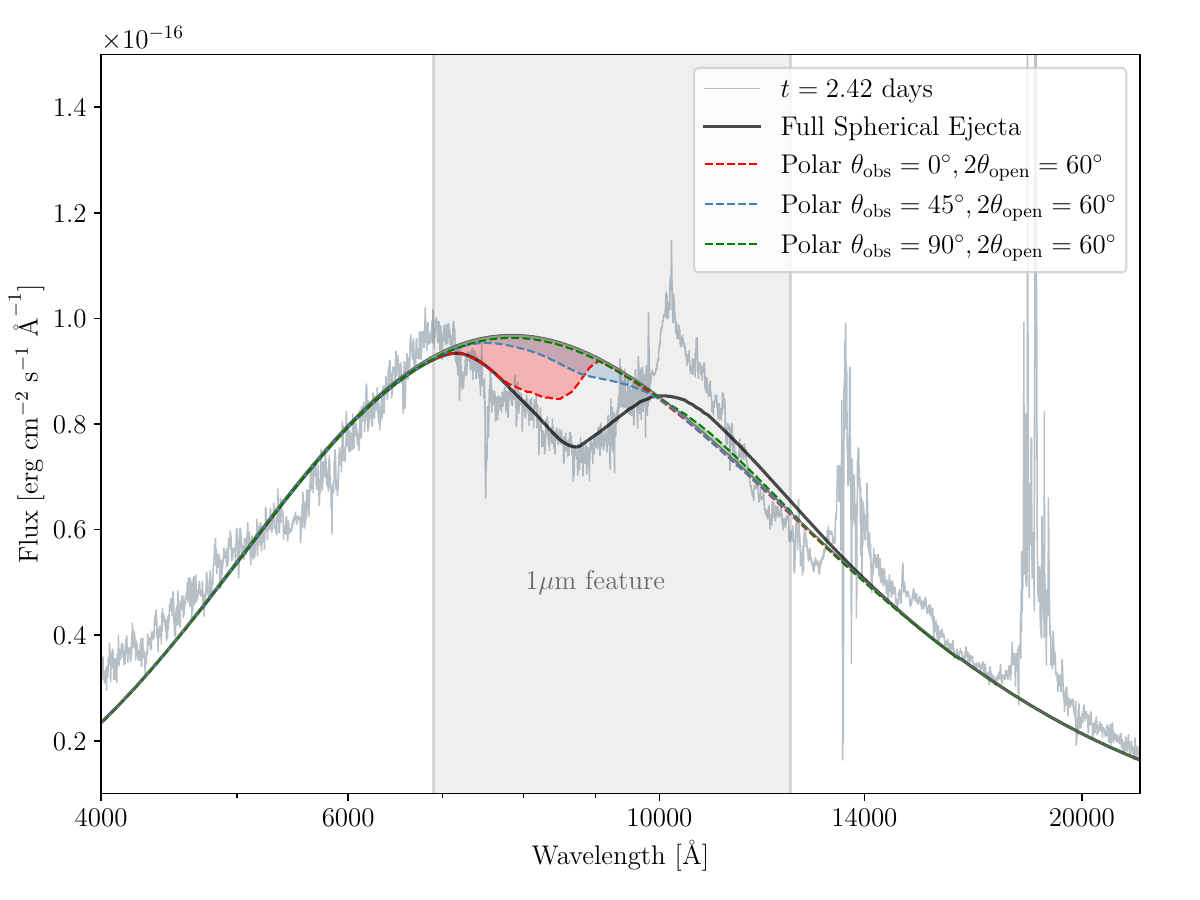}
    \caption{\textbf{Left}: Schematic of the adopted polar ejecta geometry, with a full opening angle of $2\theta_{\rm{open}} = 60^\circ$. Helium as the representative species is confined to the blue cones; the equatorial region contains no helium.
\textbf{Right}: Spectrum calculations for different observer inclinations. For an on-axis, $\theta_{\rm{inc}}=0$ ejecta, only the highest velocity absorption can form, without any emission contribution. For $\theta_{\rm{inc}}=90^\circ$, mostly emission with little absorption can form. }
    \label{fig:polar-ejecta}
\end{figure*}

\subsection{Impact on level populations}

We note that while the ionization balance is very far from LTE, we find that our $^2$P$^\circ_{1/2}$ and $^2$P$^\circ_{3/2}$ excited state atomic level populations are also not in LTE (see Fig.~\ref{fig:level-pops}). We find this to be a consistent effect across the epochs studied here, and is due to a number of factors: (i) the collision rates are too low, and insufficient to guarantee that LTE holds (ii) the exciting radiation field is diluted, which cannot excite the level to attain population more than $W_{\rm{rel}}(v) n_{\rm{LTE}}$ (see, e.g.\ \citealt{AbbottLucy1985}), where $W_{\rm{rel}}(v)$ is the geometric dilution factor described in Section \ref{sect:radiation-field} (iii) the ejecta in the line-forming region sees the photosphere as receeding, and due to Doppler shift, the atoms in the said region see the photons at a \emph{cooler} temperature than emitted. 

\subsection{Impact of NLTE ionization and excitation on spectral lineshape}

\label{sect:NLTE-lineshape}
The shape and extent of the spectral line depends on the ionization structure, atomic level populations and radial
density profile of strontium atoms (here, strictly proportional to the mass density). Since the first two of these differ significantly between LTE and NLTE, one may expect this difference to affect the shape of the $1\,\upmu$m spectral feature.

We calculated the spectra for both LTE and NLTE models for the same velocities and radial density profile of the ejecta, and found clear differences. In Fig.~\ref{fig:lineshape-LTE-NLTE}, we show that for the LTE model the absorption extends to greater blueshifts, from the highest velocity ejecta. This is because in LTE even the outermost ejecta (up to $v_{\rm{max}} = 0.5c$) contains substantial \srii, whereas at these high velocities in our NLTE model there is negligible remaining \srii when non-thermal ionization is taken into account; therefore the opacity due to material at those velocities is insignificant (e.g.\ Fig.~\ref{fig:source-function}). Due to this, it is not only the high velocity absorption that is different in LTE vs. NLTE models, but the region where the rest-frame emission comes from also shows minor differences. Note that the strontium abundance assumed in either model is different, and is chosen such that the strength of the feature roughly matches observations.

These differences can affect derived photospheric velocities and complicate inferences about the geometry of the kilonova \citep{2023SneppenSphericalSymmetry,2024CollinsGeometry}. 
We remark that the NLTE effects on the lineshape are partly degenerate with the assumed radial density profile. Regardless, if one were to ``infer" a radial density profile of the kilonova ejecta from fitting the lineshape (modulo the uncertainties in doing so), one would require steeper radial density slopes from an LTE model
than a NLTE model taking into account non-thermal ionization and geometric dilution of the radiation field.

\subsection{Impact on abundance inferences}

Inferences on the true abundance of an element in the kilonova ejecta based on the strength of line features relies on the calculation of the ionization and the level populations both of which substantially differ between NLTE and LTE. The combined effect of these two is such that any estimate of the strontium abundance derived from LTE calculations can be orders of magnitude off in the inferred mass fraction of strontium in the ejecta. And indeed, \cite{2019NaturWatson} required just $\sim 10^{-5}\ \rm{M}_{\odot}$ of strontium to produce the line feature in their LTE work. While the masses we require are similar at $1.43$\,days, they start to diverge considerably at later epochs. Given the uncertainty in the radioactive heating rate $\dot{Q}_{\rm{\beta}}$ of \rproc material \citep{Barnes2021NuclearUncertainties,Kullman2023NuclearUncertainties}, the thermalization efficiency $f_\beta$, and the density distribution, inferring the true abundance of any element in kilonovae remains difficult even with NLTE physics.


\section{Geometry and spectral lineshape: Influence of spatial confinement}

\label{sect:geometry-inclination}

The models we considered so far assumed that the kilonova ejecta is spherically symmetric and the line-forming species is distributed uniformly through this sphere. Numerical simulations of binary neutron star mergers (e.g.\ \citealt{Just2023EndToEnd,CombiSiegel2023,Kiuchi2023ShortLived}, to name a few) predict mass ejection that deviates from spherical symmetry, with a wide distribution of electron fraction ($Y_e$) within the ejecta resulting in nucleosynthesis with asymmetric elemental distribution. Certain species might also be synthesized predominantly in specific parts of the ejecta. For instance, in a lot of recent hydrodynamical simulations, helium is confined to the polar ejecta (e.g.\ see Fig.~5 of \citealt{Sneppen2024Helium2}) where neutrino-driven winds raise the $Y_e$ giving the conditions for $\alpha$-rich freezeout to cause bulk helium production.

Line formation in expanding atmospheres depends greatly on the geometry of the emitting surface. For asymmetric ejecta, the observed spectrum also depends on the observer inclination (e.g. \citealt{Shingles2023,groenewegen20252dendtoendmodelingkilonovae}). Under certain geometries and inclination angles, the species of interest may not be able to produce absorption or emission at all.

To this end, here we investigate in such geometries whether strontium or helium would be able to produce the observed $1\,\upmu$m spectral line. The peanut shaped geometry of helium distribution in the polar ejecta (see \citealt{Sneppen2024Helium2}, Fig.~5) can be approximated as a biconical hourglass. Note that while our calculation is done for helium, the results are completely general for any element including strontium which has that geometric confinement.

We note that detailed investigations of the spectral shape of the $1\,\upmu$m feature, both for the observed spectrum of AT2017gfo \citep{2023SneppenSphericalSymmetry} and of ejecta from merger simulations \citep{2024CollinsGeometry} have been done before. However, the parameterization in those studies only considered whether the observed spectral line profile requires an ellipsoidal geometry of the ejecta, or if it is consistent with spherical symmetry.

\subsection{Polar ejecta line calculations}

We assume a polar ejecta represented by a biconical hourglass Fig.~\ref{fig:polar-ejecta}), with some inclination of $\theta_{\rm{obs}}$ to the observer's line of sight. Visual inspection of output from hydrodynamical simulations shows that the polar ejecta has an opening angle $2\theta_{\rm{open}} \sim 60^\circ-90^\circ$. We assume that the species of interest (helium or strontium) is present only inside these blue cones, and not present in the equatorial regions. Therefore, they will not have any optical depth outside the cones. The photosphere, as described in Section~\ref{sect:radiation-field}, is still assumed to be spherically symmetric and therefore the prescription for the illuminating radiation field remains the same as before and the entire NLTE calculation can be carried forward in the same fashion.\footnote{We point out a minor caveat that in cases if He comprised a large fraction of the ejecta by mass in that region, the local radioactive heating rate $\dot{Q}_\beta$ would be reduced since He does not contribute to any $\beta$-decay.}  

In the right panel of Fig.~\ref{fig:polar-ejecta}, we show calculations of the spectrum for a fixed abundance $X_{\rm{He}}$ and fixed opening angle $2\theta_{\rm{open}}$ but varying inclination angles. We find that if the polar ejecta is completely on-axis with the observer ($\theta_{\rm{obs}} = 0$) and the opening angle is $2\theta_{\rm{open}} = 60^\circ$, the absorbing species will be able to produce the most blueshifted absorption corresponding to the highest projected velocity of the ejecta, but not the low velocity absorption or any emission. 
As the inclination angle is increased ($\theta_{\rm{obs}} \sim 45^\circ$) this polar ejecta covers a wider range of velocities projected along the line of sight of the observer, and can now produce the entire range of absorption similar to the spherically symmetric case. But since this narrow cone blocks a smaller fraction of the emitted radiation, the strength of the absorption becomes weaker, and one requires a larger abundance of the species to match the observed feature. Finally, for $\theta_{\rm{obs}} = 90^\circ$ there is hardly material in the region where absorption forms, but can contribute to emission. Note that the strength of the emission is still suppressed relative to the spherically symmetric case, because a $2\theta_{\rm{open}} = 60^\circ$ cone covers only a third of the solid angle visible to the observer, which reduces the total number of photons scattered to the observer's line of sight by a factor of three.

\subsection{Implications for polar confined species in AT2017gfo}

\begin{figure}[!h]
    \includegraphics[width=0.49\textwidth]{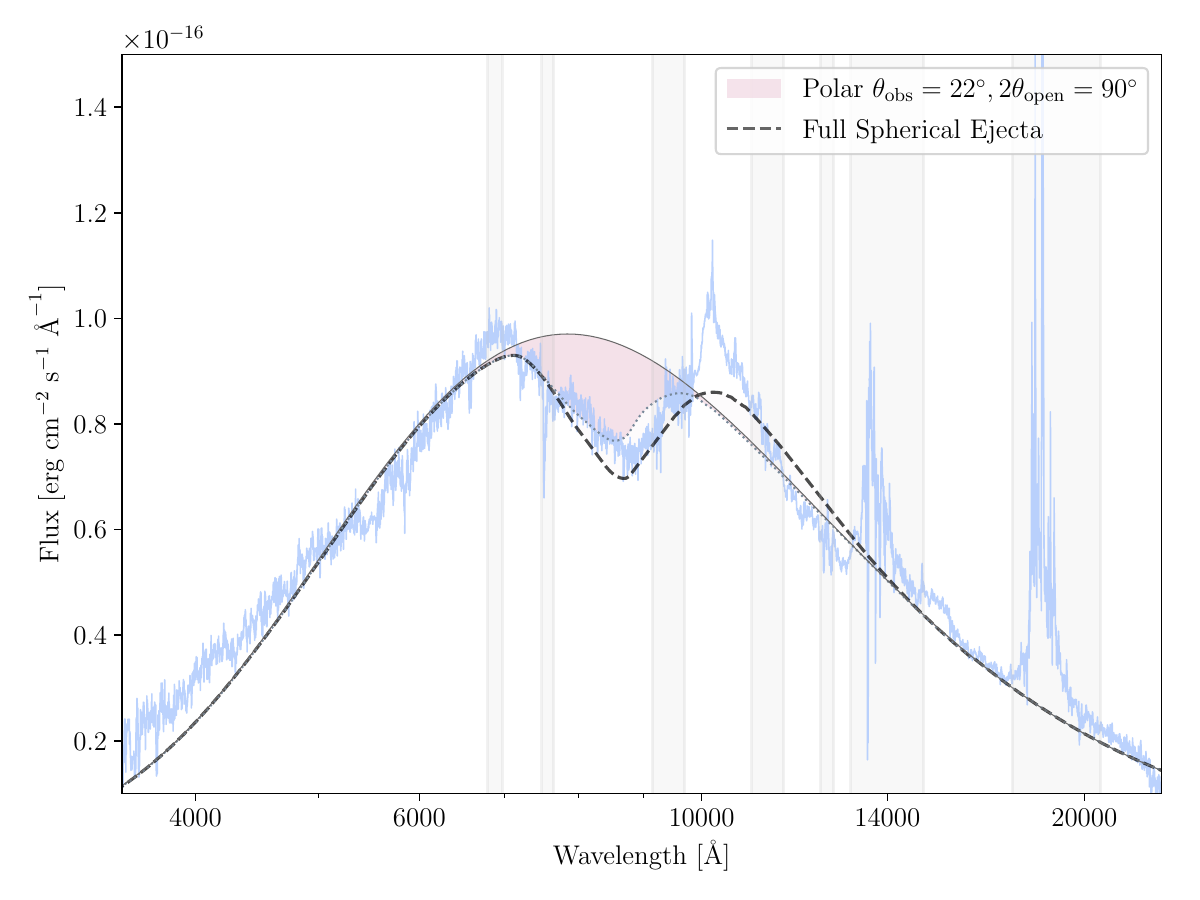}
    \caption{Comparison of synthetic \hei 1083.3 nm line profiles with $t=2.42$\,days X-Shooter spectrum for two cases: (1) helium confined to the polar ejecta (pink shaded) and (2) helium distributed spherically (black dashed line). For the AT2017gfo inclination (22$^\circ \pm 3^\circ$; \citealt{2022MooleyNature}), the polar configuration can still produce the entire absorption but little or no red emission.}
    \label{fig:polar-good-match}
\end{figure}

Fortunately, the observer inclination for AT2017gfo's polar axis is known from the afterglow emission left by the relativistic jet \citep{2022MooleyNature} to be $22^\circ \pm 3^\circ$. Therefore, we can fix $\theta_{\rm{obs}} = 22^\circ$, in which case we find that if helium were confined entirely to the polar ejecta, the inclination is such that it would be able to form almost the entire range of absorption (see Fig.~\ref{fig:polar-good-match}), but none of the observed emission. A good match to the entire absorption trough can be obtained if $2\theta_{\rm{open}}$ is increased to $90^\circ$.

If the observed $1\,\upmu$m feature is a true P\,Cygni, it requires material to be distributed in a spherically symmetric fashion to explain the observations. However, what constitutes ``emission" depends on the assumption of where the true continuum lies, over which emission forms. Our adopted continuum in these calculations is motivated by the fact that this choice recovers a luminosity distance of AT2017gfo consistent with values known from independent methods (see \citealt{2023SneppenHubbleConstant}). 

Regardless, even if were the case that in AT2017gfo any bulk production of helium was confined to the poles, geometry alone does not rule out its contribution to the absorption in the spectrum. But a different species could be required to explain the observed emission. Addressing the strontium vs. helium question requires additional considerations, which we discuss in the next section.



\section{Strontium vs.\ helium: Which element is responsible for the feature?}

\label{sect:sr-vs-he}

\begin{figure}[!h]
    \centering
    \includegraphics[width=\linewidth]{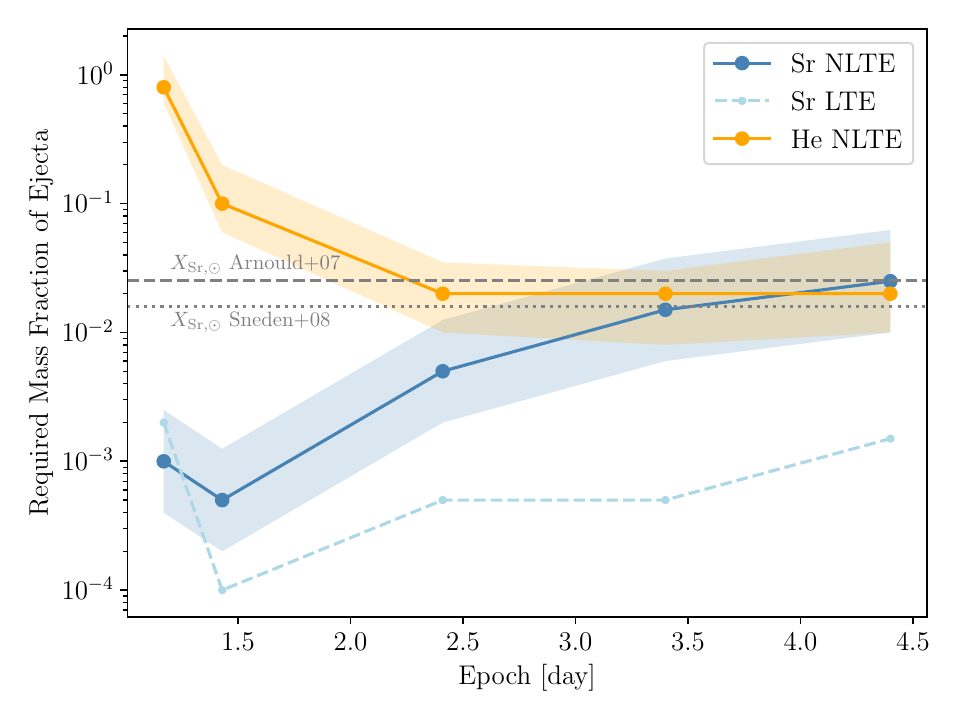}
    \caption{The abundance of strontium and helium required to explain the $1\,\upmu$m feature at different epochs, for a uniform abundance model. The dashed and dotted gray lines indicate the $X_{\rm{Sr}}$ in the solar \rproc patterns of \cite{Arnould2007} and \cite{Sneden2008} respectively.
    Clearly, the $1\,\upmu$m feature can be formed by very little strontium at early times, around when it emerges. At that same time, the whole ejecta's mass would have to be helium to explain the feature with $1083.3$\,nm \hei line. Over time, the required abundance of the Sr increases and that of He decreases. The LTE Sr model consistently requires less abundance than the NLTE model.}
    \label{fig:req_mass_fractions}
\end{figure}

In the previous sections we showed the feasibility of the two elements in being able to form the spectral feature. It remains important to address which of the two elements (or both) is responsible for it. In Fig.~\ref{fig:req_mass_fractions}, we compare the mass fractions of strontium and helium required to individually account for the observed $0.8{-}1\,\upmu$m absorption. The calculation assumed spherical symmetry and the same radial density profile for both species. 

We included two sources of uncertainty in the estimates \textemdash the radioactive heating rate, and the photospheric radiation temperature. 
While predictions for the radioactive heating rate $\dot{Q}_\beta$ differ substantially across nuclear mass models \citep{Barnes2021NuclearUncertainties}, the difference is rather small if only the newer models are considered (see \citealt{Kullman2023NuclearUncertainties}, Fig.~22). However, beyond the nuclear physics uncertainty, the heating rate depends on the ejecta's electron fraction and specific entropy, leading to differences of a factor of few (e.g.\ \citealt{Just2023EndToEnd}, Fig.~3(i)).

Therefore, we adopted $\dot{Q}_\beta$ being a factor of two larger or smaller, and a temperature uncertainty of $\pm 200$\,K. The resulting abundance uncertainty is larger for strontium, due to its higher sensitivity to non-thermal ionization. For helium, the main uncertainty comes from temperature sensitivity of photoionization.

A striking characteristic of these two elements is that the abundance of them required individually to explain the absorption evolves with epoch, and their trends go in opposite directions.  Around $t\approx1.2{-}1.5$\,days, even $0.1\%$ strontium by mass can alone explain the feature. At these epochs it is very difficult to suppress the \srii lines even if we assume higher radioactive heating rate $\dot{Q}_\beta$. But the required abundance of strontium increases with each epoch. For helium, the mass of helium needed to explain the feature at $1.17$\,days is comparable to the ejecta mass ($X_{\rm{He}} \sim 80\%$). Meanwhile, at $2.42$ days, $2\%$ helium by mass would be able to account for the observed feature.

If the high amount of helium required at $t=1.17$\,days were the true abundance of He in the ejecta, this amount of helium would overproduce the feature at $t\gtrsim 2.4$ days, as was argued also by \cite{Sneppen2024Helium1}. Meanwhile, if we assume that the $X_{\rm{He}}\sim 2{-}3\%$ He required at later epochs is the true abundance, it would have negligible contribution to the feature at $1.17$\, days. We note that due to its small atomic mass, $2\%$ helium by mass fraction is still $\approx 40\%$ by number density, and therefore is not a small abundance.





It is then quite clear that at least when the feature first emerges, it cannot come from helium, but can be easily produced by strontium. Second, as we argued in Section~\ref{subsect:emergence-1um}, the time at which the feature emerges is consistent with a strontium interpretation. At later epochs, helium can still contribute to the feature, and its contribution could even start to dominate over strontium. 

An increase in the required abundance of strontium, especially between $1.4$ and $2.4$\,days has been seen in some LTE studies \citep{Gillanders2022photospheric}. 
For strontium to solely account for the feature at all epochs, one would require an abundance that is radially stratified with a steep slope with velocity. A radially stratified abundance will likely affect the shape of the spectral line. One should remember that, the line forming region on different epochs is not entirely disjoint, but rather has significant overlaps. With that in mind, whether radial stratification can solve the problem can only be answered with a model that uses such an abundance distribution and explains both the shape and extent of the spectral feature, from $1.17$ to $4.40$\,days. Such a calculation is beyond the scope of the present study. 

We also point out that at $t\gtrsim 2.4$\,days our NLTE strontium model consistently requires more total strontium than LTE models, both in our own calculations (see Fig.~\ref{fig:req_mass_fractions}), and when compared to \citealt{2019NaturWatson,Gillanders2022photospheric, 2023Vieira,2024Vieira}. From a theoretical standpoint, strontium in the solar \rproc pattern has a mass fraction of $1.6{-}2.5\%$ in the \cite{Sneden2008,Arnould2007} patterns.  Hydrodynamical simulations of binary neutron star mergers predict strontium abundances that are either approximately solar \citep{Just2023EndToEnd,2025Bernuzzi} or can even be overproduced depending on the ejecta conditions \citep{2022Perego}. Therefore, even the $X_{\rm{Sr}} = 2.5\%$ abundance required at $t=4.40$\,days in our model is still quite consistent with \rproc expectations. As for helium, the abundance is predicted to be as high as $\sim 40\%$ in case of long-lived merger remnants \citep{2025Bernuzzi,Sneppen2024Helium2}. The $X_{\rm{He}} \lesssim 5\%$ upper limit on abundance of helium at $4.4$ days is well-within the expectations even for the case of mergers with short-lived remnants \citep{Sneppen2024Helium2}.

Also, in absolute terms, the abundances of strontium quoted here should only be treated as rough estimates, as these numbers are degenerate with the assumed radial density profile (see Fig.~\ref{fig:density-lineshape}). A steeper power law slope in the inner ejecta would leave less mass in the outer ejecta, and one would require a higher $X_{\rm{Sr}}$ to form the line with the same strength. It is only the column density, or the mass of strontium contained in the line-forming region, that is constrained. But it should be noted that changes to the density profile will affect both $X_{\rm{Sr}}$ and $X_{\rm{He}}$ in the same way. 


\section{Summary \& Conclusions}
\label{sect:summary}

An accurate identification of which element in the ejecta contributes to which line feature is important for identifying signposts of \rproc nucleosynthesis in the kilonova, and establishing whether neutron star mergers are major sites of \rproc nucleosynthesis. We showed that strontium lines are very difficult to suppress, as we show that even when the ionization balance largely disfavours \srii, the minority \srii is sufficient to produce spectral features of the $1\,\upmu$m triplet and $400$\,nm resonance doublet lines. 


Throughout this study, we have emphasized the importance of NLTE physics. At the earliest epochs when the radiation field controls the plasma's ionization state, its evolution tracks LTE predictions. This changes after non-thermal ionization overtakes photoionization. Similar to \cite{Tarumi2023, 2025BrethauerIonization}, we find that the ejecta consists of multiple ionization states at all times, from $0.92$\,days to $4.40$\,days. Apart from the ionization state, the level populations are below LTE due to the dilution of the radiation field. Inferring element abundances using LTE therefore results in an underestimation by up to a dex in some cases.

The prediction of LTE that a recombination episode makes the $0.8{-}1.0\upmu$m absorption emerge between $0.92-1.17$\,days is reproduced in our NLTE models. This happens because at these times the radiation field dominates the ionization rates, and the (radiation) temperature at which photoionization rates drop is roughly the time at which the Saha equation would expect recombination from \sriii $\to$ \srii. The presence of a recombination wave is however, not reflecting that the matter and radiation have thermalized. Due to the relatively weak dependence of recombination rates on temperature, it also does not strongly constrain the electron temperature to be close to the radiation temperature.

We found that the shape of the spectral line depends on the radial density profile, excitation and ionization structure. Future studies should attempt to quantify these effects, potentially constraining the ejecta's density structure. Our work highlights the importance of calculating the necessary atomic data such as photoionization and recombination rates to carry out such studies for other species, which currently exist only for a handful of heavy elements.

We showed that the shape of the spectral line also depends on the spatial distribution of the element as well. With idealized hourglass geometries, we showed that for the observer inclination of AT2017gfo, an element confined to the polar ejecta would only be able to produce absorption but not emission.

The abundance of either strontium or helium to single-handedly account for the $1\,\upmu$m feature evolves for both elements. We conclude that the $1\,\upmu$m feature is due to strontium when it emerges, but could be taken over by helium at e.g., $4.4$\,days.  It should be investigated whether a model with a fixed abundance having both strontium and helium contributions can successfully explain the feature at all epochs. It remains plausible, since the temporal evolution between $2.4{-}4.4$\,days is dependent also on the assumed radial density profile.

\section*{Data Availability}

The NLTE solver, line formation code, and analysis scripts used in this work are available at \url{https://github.com/cartilage-ftw/kilonovae}. The spectra used in this work are available on the GitHub repository of \url{https://github.com/Sneppen/Kilonova-analysis/} under \texttt{Spectral-Series-of-AT2017gfo}.

\begin{acknowledgements}
A.A. would like to thank Daniel M. Siegel, Oliver Just, Luke J. Shingles, Masaomi Tanaka, Alexander P. Ji, and Anders Jerkstrand for insightful discussions. 
The Cosmic Dawn Center (DAWN) is funded by the Danish National Research Foundation under grant DNRF140. A.A., D.W., A.S., R.D., C.P.B. and S.A.S. are funded/co-funded by the European Union (ERC, HEAVYMETAL, 101071865). Views and opinions expressed are, however, those of the authors only and do not necessarily reflect those of the European Union or the European Research
Council. Neither the European Union nor the granting authority can be held responsible for them.  D.J.D thanks the Science and Technology Facilities Council (STFC) of the UK Research and Innovation (UKRI) body for their support through his studentship (Project Code: ST/Y509504/1).

\end{acknowledgements}

%
%

\begin{appendix}

    \section{Are e$^-$ impact collisions important during the photospheric phase?}
    \label{app:collision-imp}

\begin{figure}[!h]
    \centering
    \includegraphics[width=0.95\linewidth]{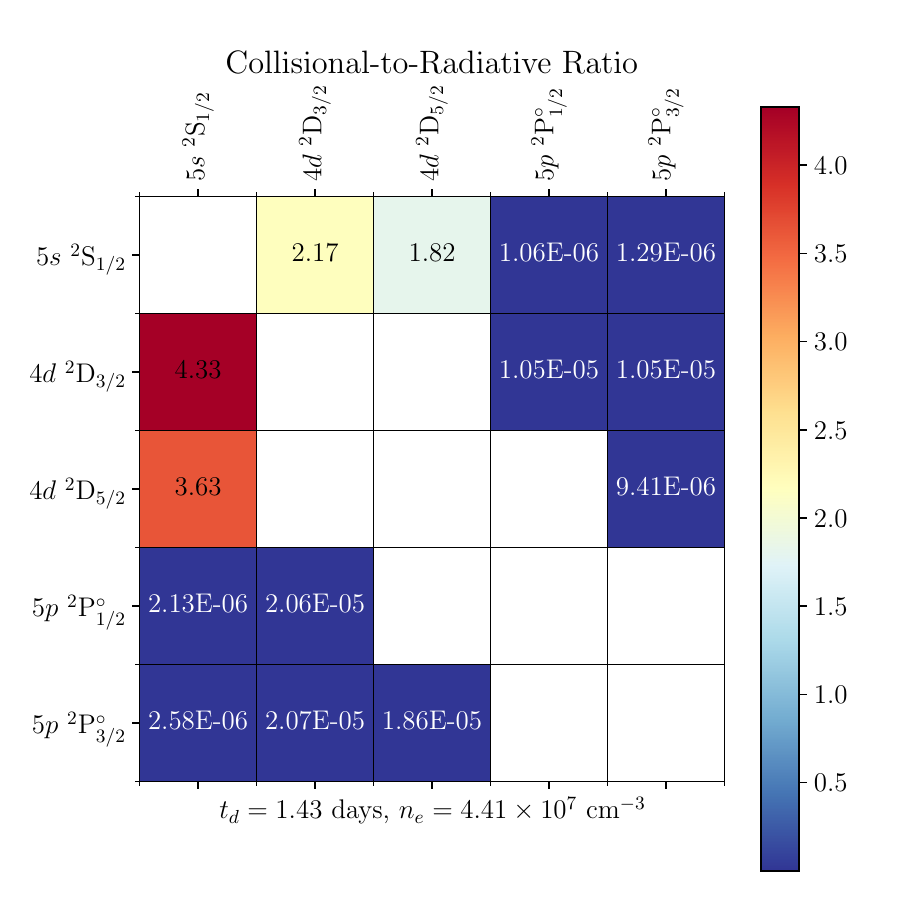}
    \caption{The ratio of collisional transition rate, to radiative transition rate of the different level transitions within \srii included here. Values larger than 1 indicate that collisions are stronger than radiation, and values smaller than 1 indicate otherwise. The matrix should be read column $\to$ row. For example, for transition from $4d$ $^2$D$_{3/2}$ $\to$ $5s$ $^2$S$_{1/2}$, the ratio is $2.17$. The lower triangle of the matrix therefore is for transitions from a lower level to an upper level, and the upper triangle is vice versa.}
    \label{fig:ratio-coll-to-radiation}
\end{figure}

    In Fig.~\ref{fig:level-pops} we showed the atomic level populations of two levels in \srii calculated with NLTE physics. This calculation was done with both radiation and electron impact collisions turned on. However, we found that the influence of the collisions in setting the level populations for the \srii levels considered here was negligible, as if collisions were completely absent. We also found that the level populations were dictated entirely by the radiation temperature, and varying the electron temperature did not change the populations.

    In cases where atomic data is limited, it is important to assess when physical processes such as electron impact collisions can be neglected. It is also important whether the fact that we do not know the true electron temperature at a given epoch (and which is expected to rise with time, see \cite{2021Hotokezaka,2022PognanSteadyState}) is going to influence spectral modelling. In the photospheric phases of kilonovae, the radiation field may be dominant in driving the transitions. This may not be true in the nebular phase, where collisions control the level populations. 
    
    Typically, it's the competition between radiation and collisions that will set the level-populations. Collisions tend to drive the level populations towards LTE. The radiation field of the kilonova photosphere is anisotropic, and cannot excite level population greater than $W_{\rm{rel}}(r,r_{min}) n_{\rm{LTE}}$ where $n_{\rm{LTE}}$ is the level population expected from a Boltzmann distribution and $W_{\rm{rel}}(r, r_{min})$ is the geometric dilution factor.
    
    One can define a critical density to quantify when electron collision rates become comparable to radiative de-excitation, as $n_{crit} = A_{ul}/q_{ul}$ where $A_{ul}$ is the Einstein coefficient for spontaneous emission of the transition, and $q_{ul}$ is the collisional rate coefficient for electron impact excitation. Typically these critical densities are $n_{crit} \sim 10^6 - 10^8$ cm$^{-3}$ for ground-state forbidden lines of \srii, and $10^{12}-10^{14}$ cm$^{-3}$ for resonance lines. But when a line is optically thick, the ``effective" radiative transition rates get reduced by a factor equal to the Sobolev escape probability $\beta_{\rm{esc}} = (1-e^{-\tau})/\tau$. Intuitively this happens because a photon that is emitted is immediately reabsorbed by another atom of the same species. Therefore, as a population, atoms of this species are not losing their excitation to emission. This reduction in effective transition rates leads to a reduction in the critical density by a factor of $\beta_{\rm{esc}}$ as well, such that an effective critical density $n_{\rm{crit,eff}} = \beta_{\rm{esc}} A_{ul}/q_{ul}$ decides whether we are in the regime where collisions are important. We find typical values of $\beta_{\rm{esc}}$ to be around $0.02-0.10$ for the $408$\,nm and $421$\,nm resonance doublet lines, and $\sim 0.5-0.8$ for the $1\,\upmu$m triplet lines.

    In the particular case of \srii studied, here we plot the ratio of net collisional rate to the radiative transition rate for each transition $i\to j$ in Fig.~\ref{fig:ratio-coll-to-radiation}. Values larger than 1 indicate that collisions are more important than radiation for those transitions.  At the densities of $\sim 10^7$ cm$^{-3}$ of the photospheric epoch of $t=1.43$\,days, it seems that collisions are able to compete with radiation only for the metastable levels, and not at all for setting the excited state populations. Note that the metastable level populations are in LTE even without collisions, at a temperature set by the radiation (see Fig.~\ref{fig:level-pops}). If we artificially increase the collision rates by $\sim 10^{5}$ such that effectively collisions compete with radiation, then the electron temperature (instead of radiation temperature) starts to set the level populations of the levels. 

    While collisions probably cannot be neglected around $10$\,days in AT2017gfo when the spectrum is nebular, in the photospheric phases this suggests that not being able to include collisions in NLTE calculation for a system like \srii may not affect results significantly. When we tested in our modelling the difference made by having collisions turned on, and turned off the result was indistinguishable. The level populations do not deviate from Fig.~\ref{fig:level-pops} if the collisions are turned off.


While the above conclusion holds for \srii, one must examine regimes where this may not be true.
For instance, in \hei we find that collisions depopulate the lower level responsible for the $587.6\,$nm. We find that the $587.6\,$nm line becomes signficantly stronger in the absence of collisions.

But \hei is also special in its atomic structure. In the absence of collisions, \hei singlet and triplet states have relatively deoupled population kinetics, as intercombination (singlet-triplet) transitions are strongly forbidden. This will not be the case for heavy elements where LS coupling doesn't hold. If we take the lanthanide element lutetium, Lu \textsc{ii} has a ground state intercombination line at $351\,$nm with an $A_{ul}$ value $\sim 10^7$ s$^{-1}$, which is as strong as its singlet-singlet or triplet-triplet lines \citep{Quinet1999LuII}.

Therefore, while collisions cannot always be ignored in the photospheric epochs, for systems like \srii they may be unimportant.

\section{Derivation of the relativistic geometric dilution factor}
\label{app:mean-intensity}

\begin{figure}[!h]
    \centering
    \includegraphics[width=0.5\textwidth]{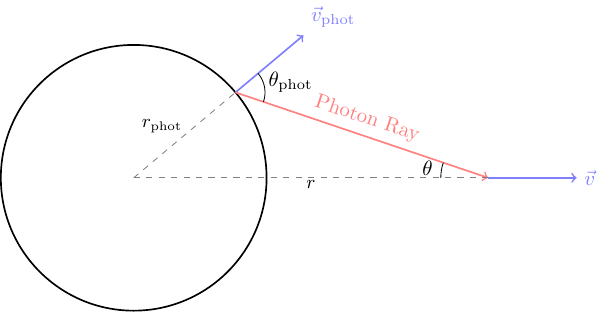}
    \caption{A fluid parcel sitting at distance $r$ from the center of the explosion sees rays from the photosphere subtending a polar angle of $\theta$, up to a maximum $\theta_c$ for which the ray is tangent to the photosphere ($\theta_{\rm{phot}}$ becomes $90^\circ$), satisfying $\sin\theta_c = r_{\rm{phot}}/r$. 
}
    \label{fig:kn-trigonometry}
\end{figure}

As described in Section~\ref{sect:radiation-field}, we assume that the photosphere emits isotropic blackbody radiation. A fluid parcel at distance $r$ from the center of the explosion does not see all the rays emitted by the photosphere. It only sees rays that lie within a cone subtended by the photosphere. This includes rays that are parallel to the observer axis with polar angle $\theta=0$, up to those which are within a cutoff angle $\theta_c$, such that $\sin\theta_c = r_{\rm{phot}}/r$ (see Fig~\ref{fig:kn-trigonometry}). Then, assuming homologous expansion ($r_{\rm{phot}}/r = v_{\rm{phot}}/v$), the angle-averaged mean intensity $J_\nu$ at radius $r$ is

$$ J_\nu = \frac{1}{4\pi} \int_{\phi=0}^{2\pi}\int_{\mu_c}^1 I_\nu(\mu) d\mu d\phi $$
where $\mu = \cos\theta$, and the lower limit of integration is the cosine of the cutoff angle $\mu_c = \cos\theta_c =  \sqrt{1 - v_{\rm{phot}}^2/v^2}$.

So far these quantities $I_\nu$ are as seen by the plasma at coordinate $v$. Because of relativistic effects, both $I_\nu$ and $\mu$ are transformed due to the Doppler effect and aberration of angles. To evaluate $J_\nu$, we need an expression in terms of what was emitted by the photosphere. We will use primed symbols for quantities that are in a frame comoving with the photosphere ($I_\nu'$, $\mu'$), and unprimed quantities ($I_\nu$, $\mu$) are as seen by the plasma at $v$. The specific intensity transforms as $I_\nu(\mu) = \delta^3(\mu) I_\nu'$, where the Doppler factor is $\delta(\mu) = [\gamma(1-\mu\beta)]^{-1}$ \citep{Castor1972}. 

To evaluate the integral above, we need to consider the transformation also of solid angles. The azimuthal angle is unaffected $d\phi' = d\phi$, while the angle cosine $\mu' = (\mu - \beta)/(1-\beta\mu)$, due to aberration of angles \citep{RybickiLightman1979,Ghisellini2013}. From the latter, we find that $d\mu = \delta^{-2}(\mu) d\mu'$.

Expressing the Doppler factor in terms of $\mu'$ yields $\delta(\mu) = \gamma(1+\beta\mu')$. If we assume azimuthal symmetry, $\int_0^{2\pi} I_\nu' d\phi= 2\pi I_\nu'$. Substituting these into our integral, the factors of $\delta$ simplify elegantly into
$$ J_\nu = \frac{1}{2} \int_{\mu_c'}^1 \gamma (1 + \beta\mu') I_\nu'(\mu') d\mu' $$
where the lower limit of the integration is adjusted due to aberration of angles $\mu'_c = (\mu_c - \beta)/(1 - \beta\mu_ c)$.

If we assume that $I_\nu'$ is isotropic (which it is for blackbody radiation), it can be taken out of the integral. We will absorb the remaining terms into a single prefactor $W_{\rm{rel}}$ such that $ J_\nu = W_{\rm{rel}} I_\nu'$, where

$$ W_{\rm{rel}} =  \frac{1}{2}\int_{\mu_c'}^1 \gamma(1 + \beta \mu')d\mu' $$


\noindent which is the geometric dilution factor corrected for relativistic effects. If we assume that $\beta$ is constant for all points on the photosphere, this integral can be evaluated analytically, which upon simplifying gives
$$ W_{\rm{rel}} = \frac{\gamma(1 - \mu_c')}{2} \left[1 + \frac{\beta(1+\mu_c')}{2}\right] $$

This is exactly what we adopted in Section~\ref{sect:radiation-field}. We caution that the $\beta$ here is not $v/c$ (where $v$ is the velocity of the fluid relative to an observer on earth), but the relative velocity of the fluid parcel at $v$ and the velocity of the photosphere $v_{\rm{phot}}$, divided by $c$. Therefore, $\beta$ is zero for a fluid parcel sitting directly at the photosphere.

One can easily verify that in the non-relativistic limit ($\beta\to 0$, $\gamma \to 1$, $\mu'_c \to \mu_c$), this smoothly recovers the classical expression $W = \frac{1}{2}\left(1 - \sqrt{1 - v_{\rm{phot}}^2/v^2}\right)$.

\end{appendix}

\end{document}